\documentclass[fleqn,usenatbib]{mnras}
\usepackage{newtxtext,newtxmath}
\usepackage{ulem,xpatch}
\usepackage[T1]{fontenc}
\DeclareRobustCommand{\VAN}[3]{#2}
\let\VANthebibliography\thebibliography
\def\thebibliography{\DeclareRobustCommand{\VAN}[3]{##3}\VANthebibliography}

\usepackage{chemformula}
\usepackage{orcidlink}
\usepackage{graphicx}	
\usepackage{amsmath}

\usepackage{color}
\usepackage{soul}

\title[CCSNe analysis with CASTOR]{Building spectral templates and reconstructing parameters for core collapse supernovae with CASTOR}

\author[A. Simongini et al.]{
Andrea Simongini$^{\orcidlink{0009-0000-3416-9865}}$,$^{1,2, 3}$\thanks{E-mail: andrea.simongini@inaf.it}
F. Ragosta$^{\orcidlink{0000-0003-2132-3610}}$,$^{4, 5}$ 
S. Piranomonte$^{\orcidlink{0000-0002-8875-5453}}$$^{1}$ and
I. Di Palma$^{\orcidlink{0000-0003-1544-8943}}$ $^{3,6,1}$
\\
$^{1}$ INAF, Osservatorio Astronomico di Roma, Via di Frascati 33, I-00078 Monteporzio Catone, Italy\\
$^{2}$ Università Tor Vergata, Dipartimento di Fisica, Via della Ricerca Scientifica 1, I-00133 Rome, Italy \\
$^{3}$ Universitá La Sapienza, Dipartimento di Fisica, Piazzale Aldo Moro 2, I-00185 Rome, Italy\\
$^{4}$ Dipartimento di Fisica “Ettore Pancini”, Università di Napoli Federico II, Via Cinthia 9, 80126 Naples, Italy \\ 
$^{5}$ INAF - Osservatorio Astronomico di Capodimonte, Via Moiariello 16, I-80131 Naples, Italy\\
$^{6}$ INFN, Sezione di Roma, 00133 Rome, Italy 
}

\date{Accepted 2024 August 5. Received 2024 July 18; in original form 2024 February 6}

\pubyear{2024}

\begin{document}
\label{firstpage}
\pagerange{\pageref{firstpage}--\pageref{lastpage}}
\maketitle
\begin{abstract}
The future of time-domain optical astronomy relies on the development of techniques and software capable of handling a rising amount of data and gradually complementing, or replacing if necessary, real observations. Next generation surveys, like the Large Synoptic Survey Telescope (LSST), will open the door to the new era of optical astrophysics, creating, at the same time, a deficiency in spectroscopic data necessary to confirm the nature of each event and to fully recover the parametric space. In this framework, we developed Core collApse Supernovae parameTers estimatOR (\texttt{CASTOR}), a novel software for data analysis. \texttt{CASTOR} combines Gaussian Process and other Machine Learning techniques to build time-series templates of synthetic spectra and to estimate parameters of core collapse supernovae for which only multi-band photometry is available. Techniques to build templates are fully data driven and non-parametric through empirical and robust models, and rely on the direct comparison with a training set of 111 core collapse supernovae from the literature. Furthermore, \texttt{CASTOR} employees the real photometric data and the reconstructed synthetic spectra of an event to estimate parameters that belong to the supernova ejecta, to the stellar progenitor and to the event itself, in a rapid and user-friendly framework. In this work we provide a demonstration of how \texttt{CASTOR} works, studying available data from SN2015ap and comparing our results with those available in literature.

\end{abstract}




\section{Introduction}
\begin{figure}
\begin{center}
\includegraphics[width=1\columnwidth]{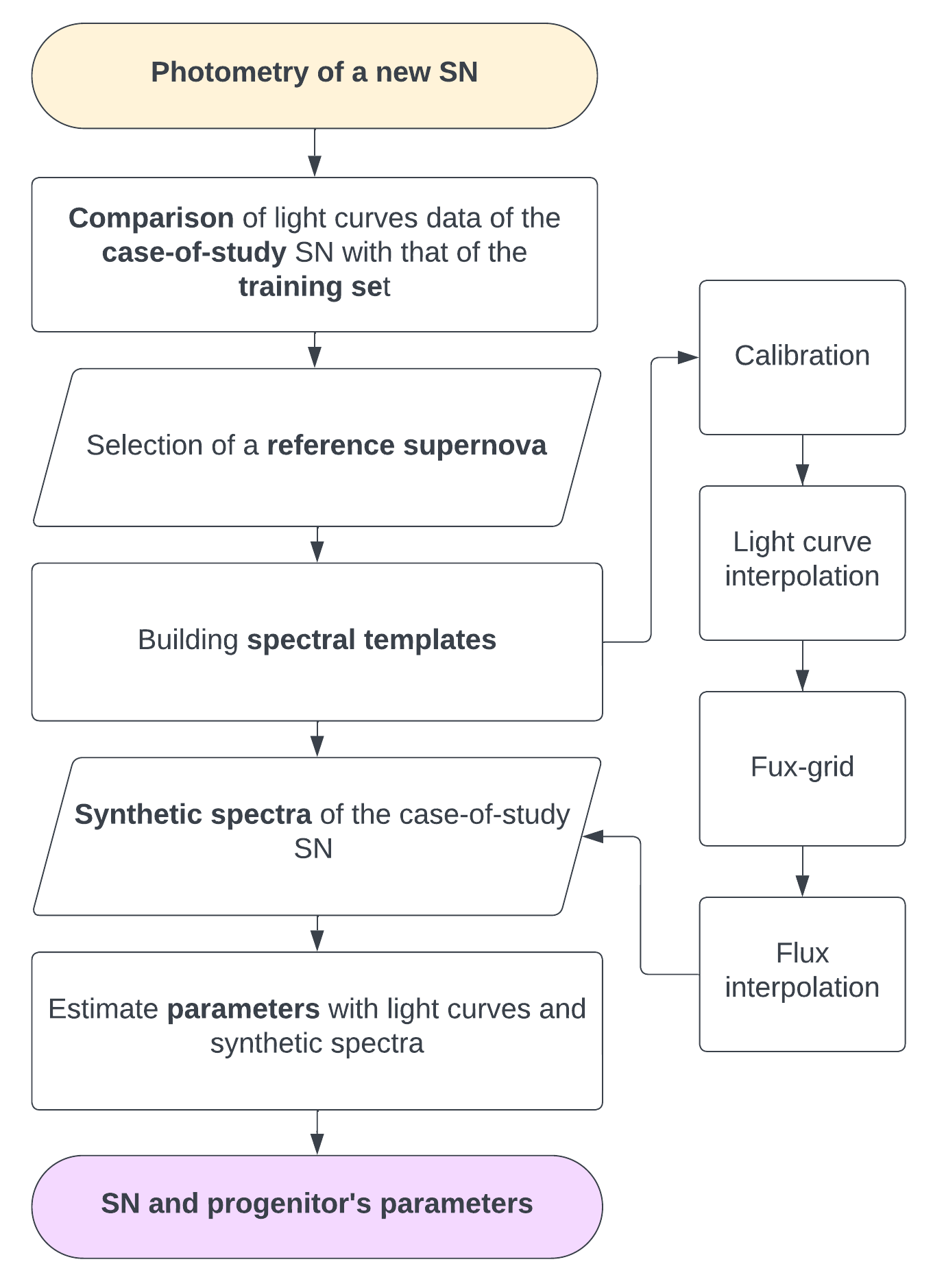}
\caption{Schematic of the operation carried out to build SN spectral templates from multi-band photometry and spectroscopy and to estimate the event parameters.}
\label{fig:flow}
\end{center}
\end{figure}

Core collapse supernovae (CCSNe) are the final evolutionary stage of a massive star ($M\ge 8 M_{\odot}$). As soon as the hydrostatic equilibrium between gravitational force pulling inward and electron-degeneracy pressure pushing outward breaks, the core collapses under its own weight. What comes next mostly depends on the initial mass of the star and can be schematized as follows: first, the core comprises itself until its density reaches a critical value ($\rho_{\mbox{\small{nuc}}} \sim 2\times 10^{14}$ g$/$cm$^3$) necessary for the strong force to intervene and halt the collapse. Then, the sudden change of velocity generates a surface of discontinuity, thus releasing a shock wave which propagates outward, interacting with the in-falling matter. The pressure behind the shock is constantly fed by neutrino interactions, leading the explosion to effectively eject the matter into the outer space \citep[and refs. therein]{Branch2017, Couch2017}. Globally, a core collapse event releases energy in the form of neutrinos (99.9$\%$ of the total energy), electromagnetic waves (0.01$\%$) and gravitational waves (0.0001$\%$), thus representing the perfect focus for current and future observational campaigns within the multimessenger field \citep[e.g.][]{Nakamura2016}.
\\ 
CCSNe progenitor stars differ to each other in terms of age, mass, environmental conditions and chemical-physical characteristics. For this reason, the taxonomy of CCSNe is complex and diverse, with significant spectral and photometric variations between different types. The standard classification scheme groups CCSNe into three main categories based on the chemical features showed in their spectra \citep[e.g.][]{Filippenko1997}. Type II supernovae (SNe II) are characterized by the presence of hydrogen absorption and emission lines and are typically linked to isolated giant or super-giant stars. Type Ib (SNe Ib) and type Ic (SNe Ic) supernovae are instead characterized by the absence of hydrogen and the presence of helium and by the absence of both elements respectively. These are theorized to arise from moderate mass interacting binaries, where a giant/super-giant star is stripped of its envelope due to mass transfer onto the companion star \citep{Nomoto1994} or from Wolf Rayet stars stripped of their envelope due to wind losses \citep{Woosley1993}. Every CCSN shows little or no quantity of silicon, which is the peculiar element in type Ia supernovae (SNe Ia). SNe Ia are not produced by the collapse of massive stars, but emerge from thermal runaway mechanisms in accreted white dwarfs binary systems \citep[and refs. therein]{Branch2017}. The differences in explosion mechanism and type of progenitor are reflected in observational properties, as the population of SNe Ia is remarkably homogeneous, both photometrically and spectroscopically, and the differences between individual objects are limited and measurable \citep[and refs. therein]{Hsiao2007}. The spectrophotometric homogeneity of SNe Ia led to the idea of creating synthetic spectral libraries (spectral templates) to substitute or complement direct observations, combining multiple observations of different objects into some representative average spectra, which are further interpolated in the time domain to produce a well sampled series of the class. CCSNe, on the contrary, exhibit great spectral and photometric heterogeneity, making the creation of spectral templates more challenging. Addressing this challenge, as part of the Supernova Photometric Classification Challenge, in 2010, \citet{Kessler2010} designed the first library of spectral templates specifically for CCSNe. That library counted 41 spectral templates based on multi-band light curves of spectroscopically confirmed CCSNe. Since the observations in ultraviolet and near infrared light were limited at that time, the templates were well-constrained only in the optical part of the spectrum. However, in the years following its first publication, the library was widely expanded thanks to the increasing number of CCSNe detected and to the employment of the The Neil Gehrels Swift Observatory, which extended the coverage to the UV part of the spectrum \citep{Roming2005}. Taking advantage of these new data, numerous methods have emerged following the first library development. Among these, an important work was that from \citet{Vincenzi2019}, who pioneered a rigorous routine to build a library of templates. Their approach relies on multi-band photometry and spectroscopy from a catalogue of extensively studied CCSNe. In a completely data-driven fashion, their routine requires the employment of light-curve fitting, dust and rest-frame corrections and cross calibration between photometry and spectroscopy. Their work aims to model the expected contamination from CCSNe events in SNe Ia cosmological surveys. 
\\ \\ 
The upcoming employment of current and new facilities within the time-domain optical astronomy, is expected to increase dramatically the number of transients detected on a daily basis. Next generation surveys and instruments like the Vera C. Rubin Observatory \citep{Ivezić2019}, the Cherenkov Telescope Array \citep[CTA,][]{acharya2017science}, the Square Kilometre Array Observatory \citep[SKAO,][]{Braun2019} and the Extremely Large Telescope \citep[ELT,][]{Gilmozzi2007}, will lead the way to intriguing discoveries at the expense of available resources necessary to spectroscopically confirm the nature of every detected object. To address this challenge and keep up with the technological advancement and the increasing volume of data, many software tools have been developed over the last two decades, especially for supernovae. Some of these tools, such as \texttt{snmachine} \citep{Lochner2016}, \texttt{RAPID} \citep{Muthukrishna2019}, \texttt{SuperNNova} \citep{Moller2020}, and \texttt{SCONE} \citep{Qu2021}, leverage machine learning and neural network techniques to classify events based on their light curves. Others, such as \texttt{SNANA} \citep{Kessler2009} and \texttt{SNCosmo} \citep{Barbary2016}, are designed for simulating light curves and spectra, focusing mostly on cosmology. Of particular impact among the community is \texttt{SuperBol} \citep{Nicholl2018}, which provides bolometric luminosity estimation and black-body fitting from a set of given light curves. The common line between these tools is their reliance primarily on light curve data or simulations, rather than real observed spectra. 
\\ \\ 
In this work we present the development and the application of a novel open-source software for data analysis called \texttt{CASTOR} \footnote{\url{https://github.com/AndreaSimongini/CASTOR}} ({Core collApse Supernovae parameTers estimatOR}). \texttt{CASTOR} takes as input light curves data of a newly discovered supernova (from now on the \textit{case-of-study supernova}) and gives as output the entire parametric map, including general parameters of the event (timing, class, absorption, distance), of the ejecta (mass, velocity, energy), of the photosphere (radius, temperature) and of the progenitor star (radius, mass). To do that, \texttt{CASTOR} employees Gaussian Process techniques, interpolating light curves and building synthetic spectral templates. This procedure follows the routine conceived by \citet{Vincenzi2019}, introducing minor changes that will be illustrated in Section \ref{sec:2} and Section \ref{sec:4}. To ensure that the synthetic spectra are well reproduced, \texttt{CASTOR} uses spectral data from a reference supernova, selected based on a Bayesian comparison of light curves from a catalogue of 111 CCSNe (from now on the \textit{training set}). Techniques to build templates are completely data-driven, while the parameters are estimated using general and common physical assumptions (i.e. spherical symmetry, partition of energy between neutrinos and photons, mass conservation). This approach is particularly significant within a multimessenger framework as it aims to extract the maximum information from electromagnetic waves, providing essential constraints that may complement linked observations made through neutrinos and gravitational waves. Furthermore, \texttt{CASTOR} can be relevant also for multiwavelength astronomy, as it rapidly analyses optical data, giving possible inputs for triggering interesting sources or constraining parameters for multi-band analysis (\textcolor{blue}{LST Collaboration 2024, in prep}). 
\\ \\ 
The paper is laid out as follows. We present a detailed description of the statistical methods employed to build spectral templates in Section \ref{sec:2}, while in Section \ref{sec:3} we describe the parameters estimation. In Section \ref{sec:4} we introduce the training set of 111 CCSNe, highlight their general properties and derive the calibration parameters. Then, in Section \ref{sec:5} we show a direct application of \texttt{CASTOR} on SN2015ap, comparing our results with those already published. Finally, we draw the conclusions in Section \ref{sec:6}. A schematic of the operations carried out in \texttt{CASTOR} is shown in Fig.~\ref{fig:flow}.

\section{Building spectral templates}\label{sec:2}

In this section, we present the statistical methods developed to build time-series spectral templates solely from the combination of the photometric observations of a new CCSN and the spectral data of a reference one. Building templates is completely data-driven and non-parametric, thus being perfectly suited for facing the great heterogeneity of CCSNe data. \texttt{CASTOR} makes extensive use of Gaussian Process (GP) regression techniques to model light curves and to smoothly interpolate spectra. A Gaussian Process is a widely employed stochastic non-parametric and data-driven interpolation technique \citep{Rasmussen2006, Kim2013, Ebden2015, Inserra2018, Saunders2018, Angus2019, Yao2020, Alves2022}. It is formally defined as a "collection of random variables, any finite number of which have (consistent) joint Gaussian distributions" \citep{Rasmussen2004}. In practice, a GP generates the best function to interpolate a given data distribution on a continuum domain. Similarly to how a multivariate Gaussian distribution is defined by its mean vector and a covariance matrix, the GP is fully defined by its mean function and its covariance function. We will write:
\begin{equation}
    f \sim \mbox{GP}(m, k),
\end{equation}
meaning that the function $f$ is distributed as a GP with mean function $m$ and covariance function $k$, also called kernel.
\\ \\ 
To perform the GP regression on light curves and spectra we use the python packages \texttt{george} \citep{Ambikasaran2015} and \texttt{scikit-learn} \citep{Pedregosa2011} respectively. The former is less complex and faster, and works more efficiently with low size datasets,  while the latter handles more precisely big sized datasets. In both cases we use a Matérn kernel function with $\nu$ = 1.5 of the form: 
\begin{equation}
    k(x, x') = A\left(1+ \frac{\sqrt{3}(x-x')}{\sigma}\right)\mbox{exp}\left(-\frac{\sqrt{3}(x-x')}{\sigma}\right),
\label{eq:kernel}
\end{equation}
where $(x-x')$ is the Euclidean distance between two measurements, $A$ is an amplitude factor and $\sigma$ is the lengthscale at which the correlations are significant. Formally, GPs can interpolate data points in more than one domain (i.e. time and wavelength when interpolating spectra). In that case, the hyper-parameters $A$ and $\sigma$ need to be defined as vectors, in order to contain the information of each domain. Moreover, due to the functional nature of the kernel, it is possible to define a composite kernel function $\bar k$ by simply adding up different kernels $k_i$: 
\begin{equation}
    \bar k(x, x') = \sum_i k_i.
\end{equation} 
This combination is particularly effective to interpolate data distributions which are irregularly sampled. To do so, we combined three Matérn kernels by maintaining the amplitude factor $A$ constant and adjusting the lengthscale $\sigma$. Note that we always set the amplitude as the average value of magnitude/flux points of a given light curve/spectrum. Moreover, when working with light curves, we set three different time-lengthscales for a kernel mixture: the medium, the maximum and the minimum distance between two consecutive points. Alternatively, when working with spectra, we set two different time-lengthscales, corresponding to the minimum and maximum sampling step (i.e. the days between two consecutive spectra) and a constant wavelength-lengthscale to a value of 70 \AA. Results for a one-dimensional Gaussian Process applied on the light curves of SN2015ap are shown in Fig.~\ref{fig:gps}. \\ \\ 
The first step to build synthetic templates is to select multi-band photometry of the case-of-study supernova. We compare each of its light curves with those of each supernova out of the training set to identify the SN that most closely resembles the photometric properties of the case-of-study (from now on, the \textit{reference supernova}). This procedure is carried out applying the chi-squared test \citep{Pearson1900}, which requires the estimation of the $\chi^2$ parameter, given by:
\begin{equation}
    \chi^2 = \sum_i^n \frac{\left(x_i - y_i\right)^2}{y_i}, 
\end{equation} 
where $x_i$ and $y_i$ are the data points of the observed and reference dataset respectively. In order to ensure comparability of time scales, we rescale each light curve according to the relative time of the explosion. Furthermore, we interpolate the multi-band photometry of the training set using the GPs at the same epochs at which light curves of the case-of-study SN are observed. We then estimate the normalized $\chi^2$ filter by filter by means of the \texttt{chisquare} module from the \texttt{SciPy} \citep{Virtanen2020} library. Consequently, we sum up each individual normalized $\chi^2$ value across all available filters and normalize the final value to the number of filters. This process results in a single chi-squared value for each supernova, normalized by the number of data points and the number of filters. Finally, the supernova with the lowest normalized chi-squared is identified as the reference supernova. 
Then, \texttt{CASTOR} has everything it needs to build time-series of spectral templates for the case-of-study supernova: light curves from the case-of-study SN and spectra from the reference SN. As previously stated, this process relies only on GP regression techniques, therefore it is completely data-driven and non-parametric. We first interpolate multi-band photometry using the GP regression, as shown in Fig. \ref{fig:gps}. To avoid non-physical behaviours, the interpolation is performed only on a time-series that ranges from the first to the last data points of each light curve. At each epoch of the time-series and for each filter, we extract from the interpolated light curves a magnitude point. Subsequently, as fully described in Section \ref{sec:4}, the magnitude points need to be calibrated before being converted to fluxes. This procedure requires the definition of two calibration parameters: an intercept $b$ and a slope $a$, that are used to calibrate magnitudes using a linear function. Once the magnitudes are calibrated, they are converted into flux densities $f$ by applying the general formula for magnitudes in the AB system: 

\begin{figure*}
\includegraphics[width=1\textwidth]{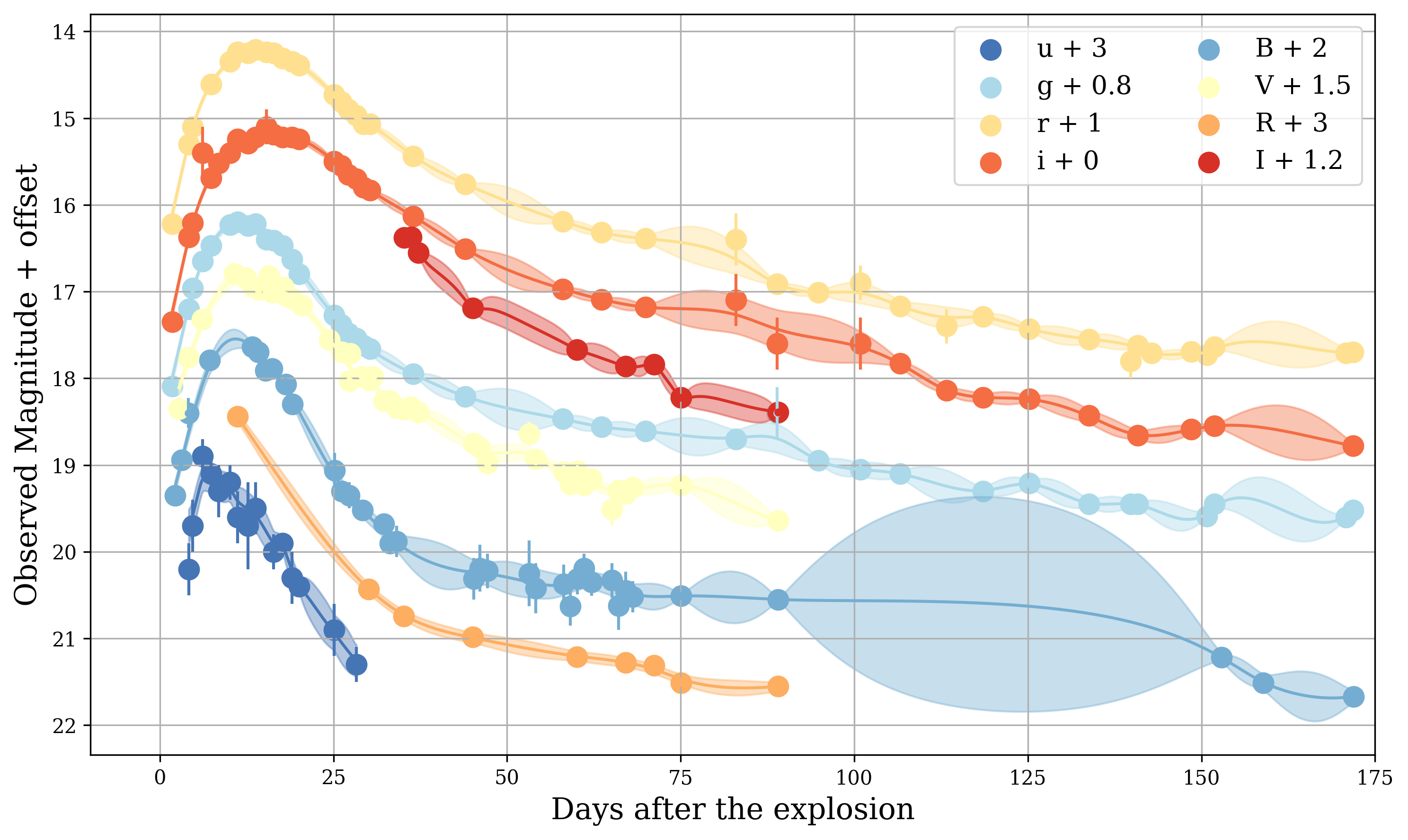}
\caption{GP interpolation of multi-band light curves for the SN Ib SN2015ap \protect\citep{Prentice2019}. We used observations in 8 filters. Data points are plotted with an offset for clarity. Filled dots indicate the published photometry. Solid lines show the results of the interpolation estimated using GPs, and the shaded area shows the uncertainty on this interpolation. While the interpolation always appear reasonable, the uncertainties are proportional to the distance of two consecutive points. As a consequence, isolated data points are connected to the bulk of points with really high uncertainties.}
\label{fig:gps}
\end{figure*}

\begin{equation}
m_{AB} = -2.5 \mbox{log}\left(f\right) - 48.60
\label{eq:magnitude}
\end{equation}
Some supernovae have been observed using the Johnson photometric system: for these cases, magnitudes are first translated into the AB system and then converted into flux densities. Every new flux point is associated to the effective wavelength of the relative filter, such that every flux is associated to a wavelength and an epoch. At this point, all fluxes, wavelengths and epochs are inserted into a two-dimensional grid. Then, to better reproduce real spectra, we add to this grid the spectral points of the reference SN, each one with its associated epoch and wavelength. This passage critically weights on the computing time, as GP regression methods struggle to handle too much data all together. For this reason, we added only part of these data, without introducing major changes in the final spectral reconstruction. Then, we interpolate flux points in the time and wavelength domains, by means of two-dimensional GP regression method on the flux grid. This technique allows the flux points to be linked across wavelengths, being at the same time constrained to the temporal evolution, and thus, to the observed spectra. The wavelength and time intervals at which interpolation is performed are dynamically defined based on the available spectral data of the reference supernova: the former is defined as the most populated range of wavelengths, while the latter is defined between the time of explosion and the last available epoch summed to the mean sampling step. Again, these precautions are taken to avoid non-physical results.

\section{Estimating parameters}\label{sec:3}

Starting from this stage of the analysis, \texttt{CASTOR} employees only the output described in Section \ref{sec:2}: the interpolated light curves and the spectral templates of the case-of-study supernova. To fully recover the parametric map of the event with no prior knowledge, no external information are taken (i.e. redshift, distance, absorption), but everything is estimated empirically from our data. In particular, \texttt{CASTOR} estimates parameters that belong to the SN ejecta and the stellar progenitor as well as it gives insights on the time of explosion, the interstellar absorption and the type of event.

\subsection{Time of explosion, of maximum luminosity and  rise}

The time of explosion of a SN is commonly taken as the midpoint in between the last "non detection" epoch and the time of photometric discovery \citep[e.g.][]{Meza2019, Bostroem2019}. The last non-detection epoch is typically set as a lower limit, defined by using independent observations of the host galaxy taken days before the discovery. However, assuming no prior knowledge of the event, we followed a different path. For every light curve $x$, we defined a last non-detection epoch $t_0^x$ by subtracting from the first observed point $t_1^x$ the average sampling step $dt^x$, given by the medium distance in time between two consecutive points. The global time of explosion is finally defined as the minimum last non-detection epoch across the entire dataset:
\begin{equation}
    t_0 = \mbox{min}\left(t_0^x\right) = \mbox{min}\left(t_1^x - dt^x\right)
\end{equation}
We take the standard deviation of all the sampling steps as its uncertainty. \\ To estimate the time of maximum luminosity $t_{max}$ we select the epoch at which the interpolated V-band light curve (alternatively the i-band) reaches its highest luminosity (i.e. its lowest apparent magnitude). Then, subtracting from the time of maximum luminosity the time of explosion, we derive the rise time $t_{rise}$. This is defined as the epoch at which the light curve becomes visible and it is typically used as the reference time of the event. 

\subsection{Interstellar absorption}

Dust and gas clouds are responsible for absorbing and scattering the electromagnetic radiation coming from an astronomical object toward an observer \citep{Trumpler1930}. The combination of these phenomena is commonly referred to as "extinction". The extinction mainly arises from the Interstellar Medium (ISM) and the circumstellar material surrounding an astronomical object, but it may also arise from the Earth atmosphere, in particular in the X, UV and IR bands. Our Galaxy also covers an important role in the overall extinction, depending on the direction of the radiation. In fact, the radiation is strongly attenuated in the denser Galactic Center, while it passes through the lighter spiral arms more freely \citep{Schlegel1998}. Since the extinction is more efficient at shorter wavelengths, some objects may appear redder then expected. This phenomenon is known as interstellar reddening \citep{Whitford1958}, and it manifests itself as a color excess. The color excess $E(B-V)$ equates as: 
\begin{equation}
    E(B-V) = (B-V)_{obs} - (B-V)_{int}
\end{equation}
where the $B-V$ terms are respectively the observed and the intrinsic color of the object. The color excess is directly related to the absorption in the $V$ band by the Cardelli's law \citep{Cardelli1989}:
\begin{equation}
    A_V = R_V E(B-V) 
\end{equation}
where $R_V$ is the reddening constant. Its value depends by the average size of the dust grains causing the extinction and can vary along different directions. The typical value of $R_V$ in our Galaxy is $3.1$ and there is no reason to think that this value would vary drastically in other galaxies. \\ 
\citet{Phillips1999} developed two empirical methods for estimating the host galaxy dust extinction for type Ia supernovae. The first method is based on a property analyzed in the work by \citet{Riess1996}: the $B-V$ colors evolution from 30 to 90 days after V maximum evolve in a nearly identical fashion. These late events, also referred to as the tail of the light curve, are dominated by nuclear decay of iron group elements. The second method, instead, uses the $B-V$ and $V-I$ colors at the peak epoch of the light curve, where the intrinsic extinction of the supernova is attenuated. This method is proposed to be used in the absence of late time events, but works less for SNe Ia due to the presence of a common envelope phase, which can generate difficult to foresee multi-wavelength features. Since CCSNe do not cross a common envelope phase, we applied the latter method. By estimating the extinction at the epoch of the peak, the total absorption depends only on the host galaxy and on the Milky Way absorption (which can be more or less effective depending on direction of the line of sight). In fact, the intrinsic extinction of the ejecta can be ignored, since the first peak represents the exit of photons from the optical thickness and the onset of the cool down phase. After the peak, recombination events with the circumstellar material are likely to happen, giving rise to a plateau in the light curves and increasing the opacity. \\ 
Operationally, we estimate the color excess $E(B-V)$ subtracting the $B$ and $V$ magnitudes taken from the interpolated light curves at $t_{max}$ \citep[alternatively, if the $B$ and $V$ bands are not available, we use the $g$ and $i$ bands, converted into $B$ and $V$ according to][]{Cardelli1989}. Then, applying the Cardelli's law with $R_V=3.1$, we compute the total absorption coefficient, which is assumed constant throughout every band. The uncertainty is taken by propagating the uncertainties of the magnitude points. The result should only depend on the host galaxy and Milky Way extinction, as the intrinsic extinction of the supernova should be close to zero. Since we assume not to know the position of the source, these two contributes are not further separated. However, a proper mapping of the source would be necessary for a more precise estimate of this parameter.

\subsection{Expansion velocity}

CCSNe spectra are characterized by different features, such as emission/absorption lines and P-Cygni profiles, which are notably the strongest and the most recognizable ones. A P-Cygni profile originates when the radiation emitted from the explosion crosses the circumstellar material moving toward the observer. As a consequence, it exhibits a broad emission line together with a blue-shifted narrow absorption line \citep[and refs. therein]{Robinson2007}. This peculiar profile is symptom of an intrinsic Doppler effect, which can be measured by estimating the relative shift from the theoretical absorption line. The Doppler shift effect equates as: 
\begin{equation}
    \frac{\lambda_{obs}-\lambda_{rest}}{\lambda_{rest}} = \frac{v_{obs}}{c}, 
    \label{eq:dop}
\end{equation}
where $\lambda_{rest}$ is the theoretical value of an absorption line and $\lambda_{obs}$ is the observed value of the same line, corresponding to the midpoint between the maximum and the minimum of the P-Cygni profile. When dealing with extra-galactic objects, one additional effect must be taken into account: the galactic recession. Observations held during the last century, showed that galaxies recede from Earth \citep{Friedmann1922}, causing the light coming from these galaxies to shift to lower frequencies. The cosmological redshift is similar to the Doppler effect as it equates in the same way defining $z=v/c$ as the redshift.
\\ \\ 
To compute the expansion velocity using the Doppler effect is a delicate task, thus we let the user take an active role in the determination of this parameter. First, we plot each synthetic spectrum with an user-selection option to allow pointing the epoch at which we want to perform the analysis of the spectra. Then, focusing on the selected spectrum, we let the user decide which element from a pre-selected list of common elements in CCSNe spectra produces the P-Cygni profile. Finally, we let the user define which intervals best contain the selected feature, intrinsically setting a prior probability on the velocity value. Once every step has been finalized, \texttt{CASTOR} will automatically estimate the evolution of the expansion velocity by fitting high degree polynomials over the selected intervals across the entire spectral time-series. To measure the goodness of each fit, we estimate the $R^2$ coefficient of determination. This is a statistical measure of how well the regression approximates the real data points. Defining $y_i$ as a data point, $\hat y_i$ as a predicted value and $\bar y$ as the mean value of the observed data, the $R^2$ value is given by:
\begin{equation}
    R^2 = 1 - \frac{\mbox{RSS}}{\mbox{TSS}} = 1- \frac{\sum_i (y_i - \hat y_i)^2}{\sum_i (y_i - \bar y)^2}
\end{equation}
where RSS is the sum of squares of residuals and TSS is the total sum of squares. We consider the fit statistically acceptable if $R^2\ge 0.9$. If this condition is verified, we estimate the velocity by applying Eq.~\ref{eq:dop} on the fitted P-Cygni profiles and we take as uncertainties the propagated uncertainty of the wavelength values. The final result will be an array of velocities.

\subsection{Distance}

The distance of the event is rather a delicate parameter to estimate, since many parameters are normalized relatively to it. To be as general as possible, without applying methods developed for specific sub-types of supernovae, we applied the Hubble's law. The Hubble's law correlate the distance of far objects with their cosmological redshift in a linear fashion \citep{Hubble1929}. This is generally expressed as: 
\begin{equation}
    d = \frac{cz}{H_0},
\end{equation}
where $d$ is the distance, $c$ is the speed of light, $z$ is the cosmological redshift and $H_0$ is the Hubble constant, which in this work is assumed equal to 70 Mpc/km/s. We measure the cosmological redshift directly from the relative shift of the theoretical position of an absorption line in the synthetic spectra with respect to the observed position. We perform the same procedures as in the case of the expansion velocity estimation, letting the user take again an active role. Note that while the velocity changes drastically as the ejecta expands, the redshift should not. For these reason, the final output of this analysis is a single, average, redshift value. Once the redshift is estimated, we apply the Hubble's law to determine the distance. Note that regardless of how accurate the estimate of the cosmological redshift is, the final distance value is intrinsically biased by two main cosmological problems, inherited by the Hubble's law itself. The first problem regards the range of validity of this law in an accelerating Universe and the second regards the uncertainty behind the precise value of the Hubble constant. 

\subsection{Spectral class}

We classify the case-of-study supernova using the information derived from the P-Cygni fitting routine, since P-Cygni profiles give immediate access to the chemical composition of the ejecta. The exhibition of strong hydrogen profiles distinguishes between type II and type Ib/Ic supernovae, which are further distinguished by the presence/absence of helium profiles. In practice, since P-Cygni profiles are directly selected by the user "by eye", the class of the event is given so. 

\subsection{Bolometric luminosity and total kinetic energy}

The bolometric luminosity is defined as the quantity of electromagnetic energy per unit of time, and is typically expressed in units of erg/s. It equates as: 
\begin{equation}
    L = 4\pi d^2 F, 
\label{eq:lum}
\end{equation}
where $d$ is the distance and $F$ is the integrated flux of the energetic source, which is obtained by integrating the flux densities taken from the interpolated light curves. The highest value assumed by the bolometric luminosity is defined as the peak luminosity $L_p$. Assuming that the energy deposited in the ejecta is completely dissipated in luminosity before the peak of emission, it is possible to further estimate the total kinetic energy of the ejecta as: 
\begin{equation}
    E_{k} = \xi L_p (t_{max} - t_0), 
\label{eq:energy}
\end{equation}
where $t_{max}$ is time of maximum luminosity and $t_0$ is the time of explosion. This equation gives direct access only to the kinetic energy that photons are able to produce. However, neutrino driven models describing the explosion mechanism of a CCSN, suggest that almost $99.9\%$ of the total energy of the explosion is released in the form of neutrinos \citep[see e.g.][]{Bethe1985, Bethe1990, Janka2017}. During the onset of the explosion, neutrinos are theorized to revitalize the shock wave by elastic scattering and neutrino capture mechanisms. Therefore the total kinetic energy of the ejecta during the first stages of evolution is strongly connected to the neutrino burst. We parameterize this relation in Eq.~\ref{eq:energy} with a factor we call $\xi$, of which value is assumed equal to $1000$, corresponding to the canonical partition of energy ($99.9\%$ to neutrinos, $0.1$ to photons) confirmed during the observations of SN1987A \citep[e.g.][]{Bionta1987, Hirata1987, Woosley1988, Fiorillo2023}.

\subsection{Mass of the ejecta}
Following the work by \citet{Arnett1982} we estimated the ejecta mass by assuming the expansion of the ejecta being homologous and the density distribution of the ejecta being homogeneous. The former is a natural consequence of a spherical shock, and the latter is made to simplify the expression of the virial theorem to: 
\begin{equation}
\frac{1}{2}M_{ej} v_{ex}^2 = \frac{5}{3} E_{k},
\label{eq:vir}
\end{equation}
where $M_{ej}$ is the ejecta mass, $v_{ex}$ is the expansion velocity, $E_{k}$ is the total kinetic energy and the numerical prefactor depends on the nature of the interaction in a uniform and self-gravitating sphere. \footnote{Note that, as commented in \citet{Wheeler2015}, there is a typo in \citet{Arnett1982} that propagated in the literature: when writing the numerical prefactor, the author reversed it. Here we use the corrected version.}

\subsection{Mass of nickel}

The late time luminosity of a supernova is expected to come primarily from $\ch{^{56}Ni}$ decays \citep[e.g.][]{Lusk2017}. \citet{Sutherland1984} simplified the difficult treatment of the propagation of $\gamma$-rays coming from iron-group nuclei decays in SN ejecta, assuming that at later times $\gamma$-rays are only subjected to one absorption. Taking advantage of this approximation, it is possible to define the rate of $\gamma$-ray energy release by radioactivity:
\begin{equation}
s = 3.90 \times 10^{10} e^{-\gamma_1 t} + 6.78\times 10^{9}\left( e^{-\gamma_2 t} - e^{-\gamma_1 t}\right),
\label{eq:s}
\end{equation}
where
\begin{equation}
    \begin{split}
\gamma_1 = 1.32\times 10^{-6} s^{-1}, \\
\gamma_1 = 1.02\times 10^{-7} s^{-1},
    \end{split}
\end{equation}
are the decay rates of $\ch{^{56}Ni}$ and $\ch{^{56}Co}$ respectively. The second term includes the $\gamma$-ray energy associated with pair annihilation, which involves the $\sim 20\%$ of the $\ch{^{56}Co}$ decays. Assuming that the late time emission is completely dominated by $\ch{^{56}Ni}$ decays, the relation between luminosity and the mass of $\ch{^{56}Ni}$ is linear:
\begin{equation}
\label{eq:nickel}
L_{Ni} = s M_{Ni},
\end{equation}
where $s$ is the energy deposit function as expressed in Eq.~\ref{eq:s}. \\ 
Following the routine from \citet{Lusk2017}, we fit a linear function on bolometric luminosity versus energy deposit points, taken from 20 days after the peak of luminosity\footnote{Note that \citet{Lusk2017} use a more strict interval taken from 120 to 200 days after the explosion. However, not every SN has photometric data after 100 days. For this reason we loosened this interval.}. To perform this fit we used the \texttt{polyfit} module from the \texttt{NumPy} \citep{Harris2020} library. The angular coefficient of the linear fit is equal to the $\ch{^{56}Ni}$ mass, while there is no priors to the intercept value, which doesn't hold a particular physical sense. The final uncertainty on the mass value is estimated as the standard error of regression.

\subsection{Photospheric temperature and radius}

Soon after the explosion, the emission of CCSNe comes from the adiabatic cooling of the outer regions \citep[e.g.][]{Eastman1996}. Due to electron scattering mechanisms, which in the highly ionized atmosphere cause the photon thermalization layer to differ from the photosphere, the emission profile of a supernova is expected to differ significantly from a pure black-body profile. \citet{Rabinak2011} quantified the departure from a pure black-body profile defining a dilution factor $\zeta$, given by: 
\begin{equation}
    \zeta = \left(\frac{T_{ph}}{T_{c}}\right)^2,
\end{equation}
where $T_c$ is the observed temperature and $T_{ph}$ is the photospheric temperature. Therefore, the emergent flux density is defined as: 
\begin{equation}
    F_\lambda = \zeta^2 \left(\frac{R_{ph}}{d}\right)^2 \pi B_\lambda(T_c),
\label{eq:flmabda}
\end{equation}
where $d$ is the distance and $B_\lambda$ is the Plank function:
\begin{equation}
B_\lambda(T_c) = \frac{2hc^2}{\lambda^5}\left(e^{\frac{hc}{k_BT_c}} - 1\right)^{-1}.
\end{equation}
To estimate the temporal evolution of the photospheric temperature and radius, we followed the same routine as \citet{Meza2019}. First, we build the spectral energy distribution (SED) using our synthetic spectra, by integrating the flux densities over every passband and associating every resulting flux point to the effective wavelength of the filter. Then, we fit it with Eq.~\ref{eq:flmabda} leaving $T_c$, $R_{ph}$ and $\zeta$ as free parameters. Wavelength data have to be further K-corrected for the cosmological redshift, such that the corrected wavelength $\lambda_c$ is given by:
\begin{equation}
    \lambda_{c} = \frac{\lambda_{obs}}{1+z}.
\end{equation}
The fit on the SED is performed using the \texttt{curve fit} module from the \texttt{SciPy} library.

\subsection{Progenitor's radius}

The radial dimension of a star can be employed to classify and distinguish between different types of stars. Indeed, giant stars have nominally up to few hundreds solar radii, while super-giants can reach up to thousands solar radii. However, this classification has not to be intended as a precise constraint, since the the radius varies significantly throughout the life cycle of a star. Methods to empirically estimate the progenitor's radius only from light curves and spectra typically rely on the assumption of specific properties (i.e. opacity, density profile) of the star \citep[e.g.][]{Rabinak2011}. However, to avoid strong assumptions, we followed another procedure. We take as progenitor's radius the photospheric radius estimated with the black-body fitting procedure at the time of explosion. The effective dimension of the progenitor star and its photosphere may or may not coincide, with the consequence that our estimate only sets an upper limit. 

\subsection{Progenitor's mass}

In order to estimate the progenitor's mass, we assumed a perfect mass conservation during the SN explosion. In this case, the progenitor's mass is equal to the sum of the mass of the ejecta and the mass of the remnant. While it is possible to directly estimate the mass of the ejecta from spectrophotometric information, the mass of the remnant is rather uncertain and requires independent observations. This is why our final estimate of the progenitor's mass $M_{pr}$ is an interval of equivalently possible values. The lower limit of this interval coincides with the sum of the mass of the ejecta $M_{ej}$ and the minimum possible mass for a neutron star to exist: 
\begin{equation}
    M_{pr} \ge M_{ej} + 1.2 M_\odot 
\end{equation}
The upper limit, on the other hand, coincides with the sum of the mass of the ejecta and the maximum mass of a black hole formed by a core-collapse mechanism:
\begin{equation}
    M_{pr} \le M_{ej} + 10 M_\odot 
\end{equation}
The uncertainties are then assumed as equals to the uncertainties of the mass of the ejecta.

\section{Training set}\label{sec:4}

\texttt{CASTOR} employees a catalogue of 111 CCSNe to select the reference supernova from. Among these, 61 are type II, 26 are type Ic and 22 are type Ib. Data of every object are collected from their relative articles or from public repositories such as \texttt{WISeREP} \citep{Yaron2012}, \texttt{VizieR} \citep[]{Ochsenbein2000} and \texttt{Open Supernova Catalog} \citep[]{Guillochon2017}. In Table \ref{tab:trainingset}, we show the entire catalogue, giving for each SN its redshift and sub-type (as reported on \texttt{WISeREP}), as well as the photometric filters and the number of light curves and spectra employed in this work. The supernovae in the training set are selected on the basis of the following criteria: 
\begin{itemize}
\item each supernova must have photometric data in at least three optical filters and one near-UV filter; 
\item each supernova must have spectral data in at least five different epochs, one within three days of the maximum peak of brightness and one after +15 days. 
\end{itemize}   
In total, we collected light curves data observed in 23 different filters, belonging to 5 of the most common photometric systems for SN observations: the Bessell system \citep[U, B, V, R, I;][]{Bessell1990}, the SDSS and the modified SDSS systems \citep[u, g, r, i, z and u', g', r', i', z';][]{Gunn1998}, the Johnson-Glass system \citep[Y, J, H, K, Ks;][]{Johnson1962, Glass1973} and the SWIFT/UVOT system \citep[w1, w2, m2;][]{Poole2008}.  \\ 
The training set is built to be representative of the diversity in SN morphology \citep{Li2011} and it plays an active role in the reconstruction of synthetic spectra, as illustrated in Section \ref{sec:2}. However, although our sample is representative, it is still not complete. For this reason, we will update our training set every time a new supernova is studied.

\begin{figure*}
\includegraphics[width=1\textwidth]{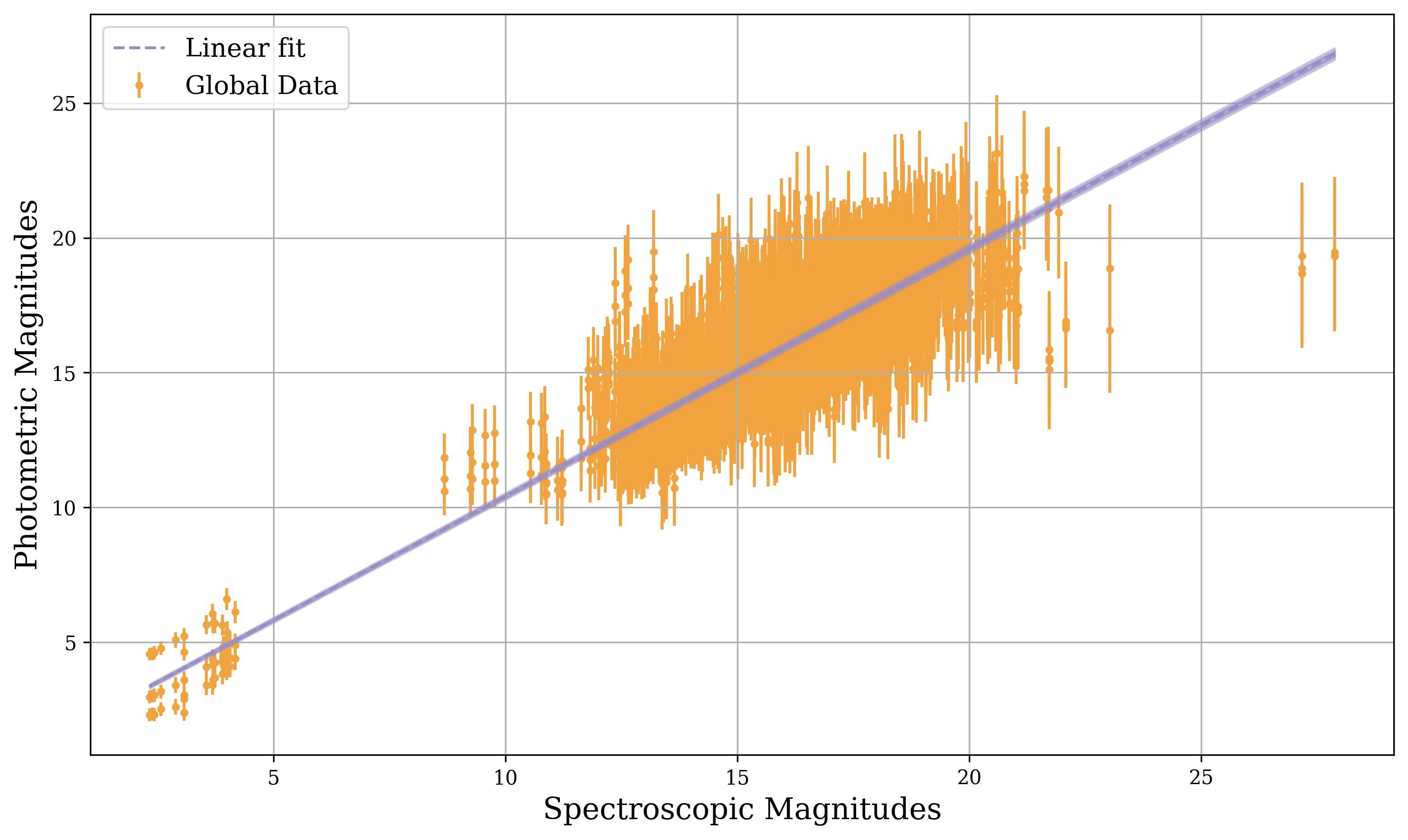}
\caption{The calibration process is carried out by fitting a linear function (purple line) on magnitudes from light curves (y-axis) and magnitudes from spectra (x-axis), taken at every available epoch and every available filter from the each supernova in the training set.}
\label{fig:mangling}
\end{figure*}
\noindent
In general, photometric magnitudes rarely match spectrometric magnitudes taken in the same bandwidth and at the same epoch, regardless of how precisely observations are performed. Despite the fact that standard reduction processes include flux calibration using a spectrophotometric standard star, there can still be differential slit and fiber losses, centring errors of the object on the slit, or non-photometric observing conditions, which may affect not only the overall normalization but also introduce a wavelength-dependent distortion on the continuum \citep{Buton2013}. Hence, when comparing magnitudes taken from light curves with magnitudes taken from spectra of the same source, observed at the same epoch and in the same filter (i.e. the same bandwidth), they don't usually match. \\ \\
We designed a calibration technique specifically for converting photometric magnitudes into spectrometric magnitudes and vice-versa. This is performed using all the data contained in the training set, therefore it aims to be the most statistically robust and thus applicable to any SN. For each supernova we first interpolate light curves by means of the GP regression technique, as illustrated in Section \ref{sec:2}. From each interpolated light curve we collect a magnitude point for every epoch and for every filter populated by spectral data. Afterward we integrate spectral flux densities over wavelength between the relative bandwidth of each filter. Finally, the integrated flux is converted into magnitudes by applying Eq.~\ref{eq:magnitude}. This process allows to insert, for every epoch and every filter, a magnitude point in a graph of photometric magnitude versus spectrometric magnitude, as shown in Fig.~\ref{fig:mangling}. We work in the magnitude space, instead of the flux space, so that the two datasets can be easily fitted with a linear function, reducing this process to the search of two parameters (the slope $a$ and the intercept $b$), which we call "\textit{calibration parameters}". In principle, if light curves and spectra were equivalent, the magnitude points would be positioned on the bisector. However, as shown in Fig.~\ref{fig:mangling}, that is not the case of real observations, proving the necessity of a calibration process. Obviously, since every point in the graph is taken from light curves and spectra which are differently calibrated, we do not expect the entire distribution to follow a completely linear fit, but only the bulk of points. We find $a = (0.919 \pm 0.003)$ and $b = (1.223\pm 0.038)$. This calibration process is similar to the more widely used "mangling" process, which is commonly employed to determine a wavelength-dependent function by comparing photometry and spectroscopy and to add data points in unpopulated regions of the spectrum by means of a smooth interpolation \citep[e.g.][]{Hsiao2007, Conley2008, Vincenzi2019, Charalampopoulos2022}.

\section{Application on SN2015ap}\label{sec:5}

\begin{figure*}
\includegraphics[width=1\textwidth]{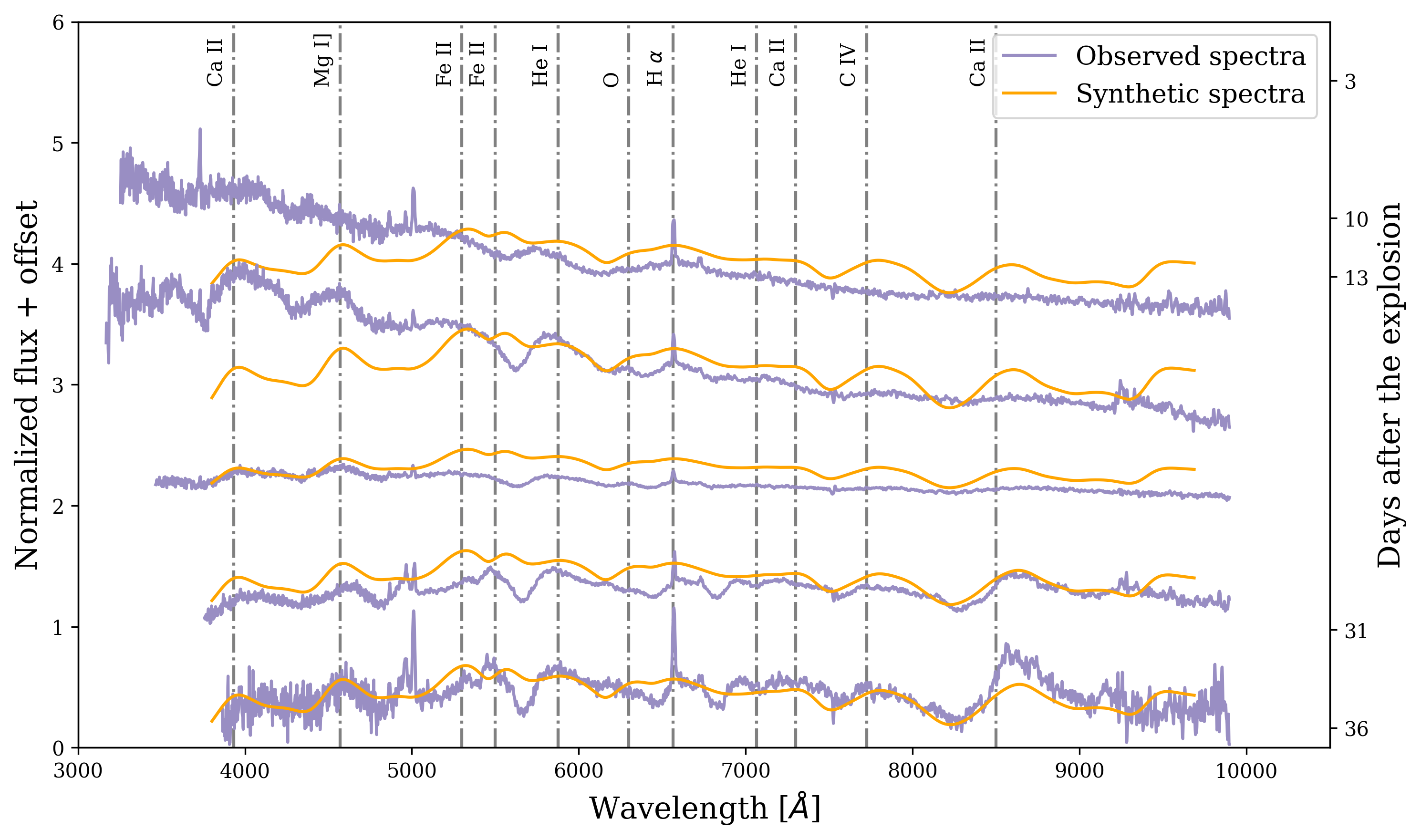}
\caption{Comparison between the observed (purple) and the synthetic (orange) spectra of SN2015ap, performed at different epochs. The time evolution has to be read from the top to the bottom. We added an offset to the normalized flux for clarity. We also show some prominent features of both datasets. The main trend of every spectrum is well reconstructed. Not all the P-Cygni profiles match the real ones, as some strong features are inherited by the spectra of the reference supernova. } 
\label{fig:comp_sp}
\end{figure*}

In order to show an application of the algorithm, we tested it on a supernova that was already studied in literature: SN2015ap. This type Ib supernova was first discovered by \citet{Ross2015}, $28''.3$ west and $16''.2$ south of the nucleus of its host galaxy IC 1776, which has a redshift of $z=0.011375\pm0.000017$ \citep{Chengalur1993}. The first observations of SN2015ap were obtained in four bands (B, V, R, I) by the KAIT, Nickel and Shane telescopes as part of the Lick Observatory Supernova Search \citep[LOSS;][]{Filippenko2001}. In addition, more observations in other four bands \citep[u, g, r, i;][]{Prentice2019} were obtained by the Las Cumbres Observatory Global Telescope Network \citep{Brown2013}. We show in Fig.~\ref{fig:gps} and in Fig.~\ref{fig:comp_lc} the light curves we employee in this work. 
\\ \\ 
From the chi-squared test, we found SN2016iae \citep{Prentice2019} as the most resembling SN out of the training set, with a normalized chi-squared of 0.003950. SN2016iae is a type Ic CCSN, discovered in the nearby galaxy NGC 1532, 17 Mpc away from Earth. We collected photometric observations in 8 bands (including the optical B, V, g, r, i) and spectrometric observations of 14 epochs. The comparison between the five available light curves from SN2015ap and SN2016iae is shown in Fig.~\ref{fig:comp_lc}. Following the procedure explained in Section \ref{sec:2}, the spectra of SN2016iae are used as constraints for the templates reconstruction. We built 50 synthetic spectra for SN2015ap, sampled one every 2.5 days from the day of the explosion to 128 days after it, accordingly to the available spectra of SN2016iae, which cover a time interval between 5 and 118 days after the explosion, with an average sampling step of 10 days. Each synthetic spectrum covers the interval of wavelength between 3802-9686 \AA. In Fig.~\ref{fig:comp_sp} we show the comparison at different epochs between synthetic and observed spectra of SN2015ap, which are collected from the \texttt{WISeREP} archive. Since we build synthetic spectra mostly from the information coming from the spectra of a different supernova and sporadic information from the main object, we do not expect the synthetic and the observed spectra to be equal. What's important is the global trend of the spectrum and the reconstruction of some P-Cygni profiles. Therefore, some lines which are not exhibited in the reference SN may not be represented in the synthetic spectrum. Finally, in Table~\ref{tab:comparison} we show the results of the parameter reconstruction. \texttt{CASTOR} required an average of 173 seconds to complete the entire analysis (not counting the time dedicated to the P-Cygni routine which, being interactive, can require also minutes). The chi-squared test and the templates building are the most time-consuming features, respectively taking an average of 53 and 116 seconds, with parallelized processes over 5 CPUs. We compare the results of our test with those from \citet{Aryan2021}. Some of their values are given without errors, making the comparison often difficult and less valuable. The methods and assumptions presented in their work often differ from those applied in \texttt{CASTOR}. Therefore, when necessary, these differences are highlighted. We found that:

\begin{figure}
\includegraphics[width=1\columnwidth]{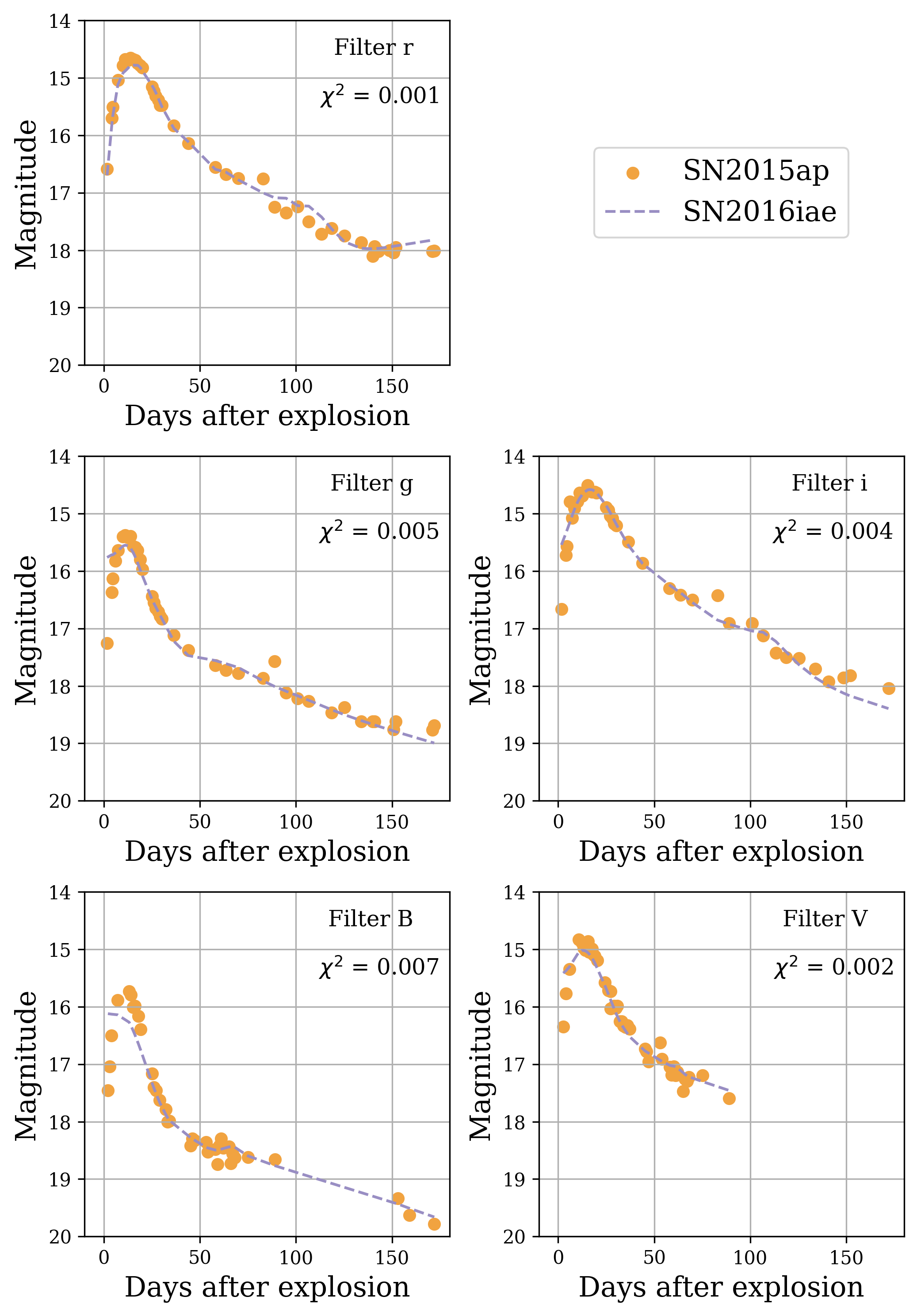}
\caption{Comparison between the light curves of SN2015ap (filed dots, in orange) and  SN2016iae (lines, in purple). The light curves of SN2016iae are interpolated by GPs regression between the same time interval of the light curves of SN2015ap to ensure time comparability. The two look very similar after the peak of maximum luminosity, while at earlier times the comparison is less valuable due to the lack of points in the SN2016iae dataset. For each comparison we highlighted the normalized chi-squared value ($\chi^2$). }
\label{fig:comp_lc}
\end{figure}

\begin{itemize}
\item in their Fig.~\textcolor{blue}{10}, \citet{Aryan2021} show the most prominent features in the observed spectra of SN2015ap. They classify it as a type Ib due to the presence of the strong characteristic He I line at 5876 \AA. They also identify unambiguous He I features at 6678 \AA \, and 7065 \AA. The spectral evolution also shows the Ca II in the near infrared and the forbidden [Ca II] feature near 7300 \AA \, at late times, with the latter possibly blended with [O II] emission at 7320 \AA \, and 7330 \AA. They also identify the Ca II H$\&$K feature near 3932 \AA \, in every spectrum, the Fe II feature near 5169 \AA, highly blended with He 5016 \AA \, and weak semi-forbidden Mg I] line at 4571 \AA. Most of these features are well reconstructed by \texttt{CASTOR}, as shown in Fig.~\ref{fig:comp_sp}. The spectral evolution shows persistent prominent features of Ca II H$\&$K at 3932 \AA \, and Ca II in the near infrared, as in the observed spectra. Moreover there are some strong features inherited by the spectra of SN2016iae: the Fe II line at 5300 \AA, the Fe II line at 5500 \AA, which seems to blend with O at 5577 \AA \, at late times and the blended O and C IV lines at 7776 \AA \, and 7724 \AA. One of the strongest features inherited by SN2016iae, which is weak in the observed spectra of SN2015ap is the semi-forbidden Mg I] line at 4571 \AA. Some He I features are present also in the synthetic spectra, but are not the most prominent ones. In particular, we can identify the He I line at 5876 \AA \, and, only at early times, the He I line at 7065 \AA. A strong P-Cygni line can be identified, especially at early times, between the H$\alpha$ line at 6563 \AA \, and the He I line at 6678 \AA. Only at late times our spectra feature the blended [Ca II] line at 7300 \AA \, and the O line at 7330 \AA \, as in the observed spectra, and the O line at 6300 \AA \, as inherited by SN2016iae. Since helium lines are not strong features in the spectra of SN2016iae, we assume that the reconstructed features in our synthetic spectra are mostly due to observed photometric data of SN2015ap. Therefore, we classify it as a type Ib. However, from our data, we could easily misclassify it as a type Ic. Despite most works classify this supernova as a type Ib \citep[e.g.][]{Prentice2019, Aryan2021}, SN2015ap was initially classified as a type Ib/c due to the exhibition of broad-lined features at early phases \citep{Tartaglia2015, Ochsenbein2000} and a recent work from \citep{Bengyat2022} re-classify it as a IIb using a novel machine-learning routine.

\item The time of explosion $t_0$, the time of maximum luminosity $t_{max}$ and the rise time $t_{rise}$ are all in good agreement, within the errors, with the referenced values. \citet{Aryan2021} determined these parameters by fitting a sixth-order polynomial, first on the R-band, then on the B-band light curves. We estimated the time of explosion applying a statistical analysis over every available light curve and determined the time of maximum by applying GPs on the V band. Finally, the rise time is estimated as the difference of these two values.

\item \citet{Aryan2021} took the Galactic extinction value from the NASA Extragalactic Database (NED) \footnote{\url{https://ned.ipac.caltech.edu/}}, assuming the host galaxy extinction as negligible. We instead applied an empirical study, determining the overall absorption as the difference between the B and V band magnitudes at maximum, and the high uncertainty comes from the propagation of GPs errors. However, due to the lack of uncertainties in the referenced extinction value, it is difficult to state the compatibility of the two values.

\item The expansion velocity result as estimated at $t_0$ is not compatible with the referenced result. However, the overall evolution of the velocity follows the expected behaviour.

\item The redshift and distance values are in agreement with the results from \citet{Aryan2021}. They extracted the redshift value directly from the host galaxy \citep{Chengalur1993} and the distance from NED, assuming a cosmological model with $H_0=73.8$ km s$^{-1}$ Mpc$^{-1}$, $\Omega_m = 0.3$ and $\Omega_\Lambda = 0.7$. We instead applied an empirical strategy, taking the redshift directly from the shift in the emission line profiles of the synthetic spectra and the distance from the Hubble law with $H_0=70$ km s$^{-1}$ Mpc$^{-1}$. 

\item The bolometric luminosity estimated at the peak is not compatible with the reference value, despite being at the same order of magnitude. \citet{Aryan2021} obtained this value by integrating the flux over a wavelength range of 4000-10000 \AA, while we estimated the luminosity directly from the interpolated light curves.

\item The total kinetic energy is in agreement, within the uncertainties, with the referenced one. At the same time, the mass of the ejecta is in agreement within 2$\sigma$, while the mass of nickel appears too low with respect to the reference value. \citet{Aryan2021} rely on a prior knowledge of opacity, following the equations proposed in \citet{Wheeler2015}. However, in the analysis proposed in this work, by using Eq.~\ref{eq:energy} and Eq.~\ref{eq:nickel}, opacity is neglected in favor of assuming complete adiabacity at the time of maximum luminosity. 

\item The temperature values are estimated at $+21$ days after the epoch of the peak luminosity. The temperature obtained in our analysis seems too low with respect to the value obtained in the referenced paper. Yet, their result is given as "$T_{BB}$", suggesting that they estimated the black body temperature without any additional correction due to a non perfect thermalization, as we made in this work. This correction is performed by the dilution factor $\zeta = (T_{ph}/T_{BB})^2$. The dilution factor at that epoch is estimated by the SED fitting routine as $\zeta=0.41\pm0.005$, resulting in a black body temperature of $T_{BB} = 6082\pm 534$K.

\item The provided values of the photospheric radius, representing their maximum values, are not compatible. In our model, a value of $\sim 45\times 10^3 R_\odot$ is strongly rejected by the boundaries we set for our fit calculation, which range between $1-20\times 10^3 R_\odot$. A potential improvement that could be implemented in the near future is to revise the prior definition, making higher values less probable but not automatically excluded.

\item \citet{Aryan2021} don't provide a clear estimate for the progenitor's radius. However when estimating the photospheric radius evolution, they provide its value at the time of explosion. In this work we assumed the photospheric radius at the time of explosion to be the upper limit for the value of the progenitor's radius. For this reason, we compared our estimate of the progenitor's radius with their photospheric radius at $t_0$. The order of magnitude is consistent and their result falls within our upper limit. 

\item \citet{Aryan2021} estimated the progenitor's mass by comparing different progenitor models with different masses with respect to the available light curves. The best matches between observations and simulations are found at $12M_\odot$ and $17M_\odot$. In this work we followed a different approach, using the empirical value of the mass of the ejecta to estimate a confidence interval with uncertainties equal to the uncertainties of the mass of the ejecta. Our interval is in agreement with their best fit, strengthening the idea of the progenitor star having either a mass of $M_{ZAMS} = 12M_\odot$ or $M_{ZAMS} = 17M_\odot$. The former would suggest a neutron star remnant, while the latter a black hole remnant, according to our prediction.
\end{itemize}
\begin{table}
\centering
\caption{The parametric map as reconstructed by \texttt{CASTOR} for SN2015ap, compared with the results from \protect\citet{Aryan2021}. Velocities are compared at the time of explosion. Photospheric temperatures are compared at $+21$ days after the maximum luminosity. The provided photospheric radius represents its maximum value.}
\label{tab:comparison}
\begin{tabular}{l|cr} 
\hline
Parameters & \texttt{CASTOR} & Aryan et al. 2021 \\
\hline
Class                & Ib                               & Ib                        \\ 
$t_0$ [MJD]          & 57272.24 $\pm$ 2.94     & 57272.72 $\pm$ 1.49       \\
$t_{max}$ [MJD]      & 57283.195 $\pm$ 2.05    & 57282.47 $\pm$ 2.56       \\
$t_{rise}$ [MJD]     & 10.96 $\pm$ 3.58        &  14.8 $\pm$ 2.2            \\ 
$A_V$ [mag]          & 0.66 $\pm$ 0.3         & 0.117                     \\
v$_{ex}$ [km/s]      & 6929.36 $\pm$ 64.95     & 9000                      \\
z                    & 0.010                            & 0.011                     \\
d [Mpc]              & 44.51 $\pm$ 1.09        & 45.07                     \\
L$_{41}$ [erg/s]     & 21.9 $\pm$ 1.07         & 35.3 $\pm$ 1.6            \\
E$_{51}$ [erg]       & 2.07 $\pm$ 0.69         & 1.05 $\pm$ 0.31  \\
M$_{ej}$/M$_{\odot}$ & 7.33 $\pm$ 2.43         & 2.20 $\pm$ 0.60           \\
M$_{Ni}$/M$_{\odot}$ & 0.06 $\pm$ 0.01         & 0.14 $\pm$ 0.02           \\
T$_{ph}$ [K]         & 3895 $\pm$ 341          & 4450                      \\
$\zeta_{ph}$ & 0.41 $\pm$ 0.005       & -                \\ 
R$_{ph}$/R$_{\odot}$ &  9015 $\pm$ 1381                          & 46982                     \\
R$_{pr}$/R$_{\odot}$ &   $<$ $8\times 10^3$      & 5948                      \\
M$_{pr}$/M$_{\odot}$ &   (8.53, 17.33) $\pm$ 2.43  & (12,17)                   \\

\hline
\end{tabular}
\end{table}

\section{Conclusions}\label{sec:6}

We have presented the development of \texttt{CASTOR} (Core collApse Supernovae parameTers estimatOR), a novel software for data analysis within the time-domain optical astronomy field. Compared to existing tools, our software represents a significant improvement in several ways. We highlight the following key features of it: 
\begin{enumerate}
\item \texttt{CASTOR} is a user-friendly and rapid software that employees solely multi-band photometric data of a newly discovered supernova to recover its entire parametric map, without the need of spectral follow-ups. Therefore it opens the window to the study of supernovae which lack of spectral information. 
\item \texttt{CASTOR} builds synthetic spectral templates by means of Gaussian Process regression techniques, following a robust routine pioneered by \citet{Vincenzi2019}. This method is non parametric and data-driven, thus it is flexible for modeling the great heterogeneity of CCSNe light curves.  
\item \texttt{CASTOR} estimates the parameter of an event with the use of empirical techniques extensively employed in literature and robust physical models. Every parameter is estimated directly from the available light curves and the synthetic spectra, with no external information about the event. According to the SN physical models assumed, every possible parameter to be extracted from electromagnetic waves is estimated. 
\item Even tough this is not its main purpose, and in this work we did not give any direct example of its usage, \texttt{CASTOR} can be easily employed to analyse supernovae using their own available spectral data, by simply defining the case-of-study supernova and the reference supernova as the same object. The results in the synthetic spectra reconstruction and the parameters estimation are expected to be more precise. An example of this different usage of \texttt{CASTOR} can be seen in \textcolor{blue}{LST Collaboration} (\textcolor{blue}{2024, in prep.}). 
\end{enumerate}
Modern and future SN surveys will gather more and more data, boosting the necessity of data analysis software like \texttt{CASTOR}. Given the large number of new photometric and spectroscopic data available every year, our training set is likely to grow, boosting our ability to recognize and characterize new events. This growth includes the addition of new spectral classes, a wider range of wavelength coverage and more diversity in redshift. As a demonstration of the software, we studied the type Ib CCSN SN2015ap. Results from this analysis show how the reconstruction of synthetic templates is capable of giving parameters which are comparable to previous estimates. However, these results can always be improved. In a future work we will make an extensive statistical analysis over a wide catalogue of CCSNe, to test the capabilities of \texttt{CASTOR} and to give robust information on the accuracy of the parameter estimation, putting particular emphasis on the class and the progenitor's mass recognition.
\\ \\ 
At the moment, \texttt{CASTOR} is employed for the comprehensive characterization of CCSNe based solely on electromagnetic information. In light of the forthcoming era of the Vera C. Rubin Observatory \citep{Ivezić2019}, this work is well-timed and aims to continue the preparatory scientific phase by focusing our attention and allocating new resources towards multimessenger and multiwavelength astronomy. In the near future, it may be possible to adapt this software to serve as a valuable tool for aggregating combined information from multimessenger observations, thereby contributing to upcoming investigations regarding the nature of core collapse supernovae, of their remnants and of their progenitors.

\section*{Acknowledgements}

This work made use of \href{https://wiserep.weizmann.ac.il}{WISeREP}, of \href{https://vizier.cds.unistra.fr/}{VizieR} and of the \href{https://sne.space}{Open Supernova Catalog}. 
IDP acknowledges the support of the  Sapienza  School  for Advanced Studies (SSAS) and the support of the Sapienza Grants No. RM120172AEF49A82 and RG12117A87956C66.

\section*{Data Availability}

The data underlying this article are publicly available in the articles listed in Table~\ref{tab:trainingset} and in their online supplementary material. The source code and the observational data compiled for this project are publicly available at the online repository \footnote{\url{https://github.com/AndreaSimongini/CASTOR}}.


\bibliographystyle{mnras}
\bibliography{biblio}

\begin{thebibliography}{}
\makeatletter
\relax
\def\mn@urlcharsother{\let\do\@makeother \do\$\do\&\do\#\do\^\do\_\do\%\do\~}
\def\mn@doi{\begingroup\mn@urlcharsother \@ifnextchar [ {\mn@doi@} {\mn@doi@[]}}
\def\mn@doi@[#1]#2{\def\@tempa{#1}\ifx\@tempa\@empty \href {http://dx.doi.org/#2} {doi:#2}\else \href {http://dx.doi.org/#2} {#1}\fi \endgroup}
\def\mn@eprint#1#2{\mn@eprint@#1:#2::\@nil}
\def\mn@eprint@arXiv#1{\href {http://arxiv.org/abs/#1} {{\tt arXiv:#1}}}
\def\mn@eprint@dblp#1{\href {http://dblp.uni-trier.de/rec/bibtex/#1.xml} {dblp:#1}}
\def\mn@eprint@#1:#2:#3:#4\@nil{\def\@tempa {#1}\def\@tempb {#2}\def\@tempc {#3}\ifx \@tempc \@empty \let \@tempc \@tempb \let \@tempb \@tempa \fi \ifx \@tempb \@empty \def\@tempb {arXiv}\fi \@ifundefined {mn@eprint@\@tempb}{\@tempb:\@tempc}{\expandafter \expandafter \csname mn@eprint@\@tempb\endcsname \expandafter{\@tempc}}}

\bibitem[\protect\citeauthoryear{Acharya et~al.,}{Acharya et~al.}{2017}]{acharya2017science}
Acharya B.~S.,  et~al., 2017, arXiv: 1709.07997

\bibitem[\protect\citeauthoryear{{Alves}, {Peiris}, {Lochner}, {McEwen}, {Allam}, {Biswas}  \& {LSST Dark Energy Science Collaboration}}{{Alves} et~al.}{2022}]{Alves2022}
{Alves} C.~S.,  {Peiris} H.~V.,  {Lochner} M.,  {McEwen} J.~D.,  {Allam} T.,  {Biswas} R.,   {LSST Dark Energy Science Collaboration} 2022, \mn@doi [\apjs] {10.3847/1538-4365/ac3479}, \href {https://ui.adsabs.harvard.edu/abs/2022ApJS..258...23A} {258, 23}

\bibitem[\protect\citeauthoryear{{Ambikasaran}, {Foreman-Mackey}, {Greengard}, {Hogg}  \& {O'Neil}}{{Ambikasaran} et~al.}{2015}]{Ambikasaran2015}
{Ambikasaran} S.,  {Foreman-Mackey} D.,  {Greengard} L.,  {Hogg} D.~W.,   {O'Neil} M.,  2015, \mn@doi [IEEE Transactions on Pattern Analysis and Machine Intelligence] {10.1109/TPAMI.2015.2448083}, \href {https://ui.adsabs.harvard.edu/abs/2015ITPAM..38..252A} {38, 252}

\bibitem[\protect\citeauthoryear{{Anderson} et~al.,}{{Anderson} et~al.}{2018}]{Anderson2018}
{Anderson} J.~P.,  et~al., 2018, \mn@doi [\aap] {10.1051/0004-6361/201833725}, \href {https://ui.adsabs.harvard.edu/abs/2018A&A...620A..67A} {620, A67}

\bibitem[\protect\citeauthoryear{{Angus} et~al.,}{{Angus} et~al.}{2019}]{Angus2019}
{Angus} C.~R.,  et~al., 2019, \mn@doi [\mnras] {10.1093/mnras/stz1321}, \href {https://ui.adsabs.harvard.edu/abs/2019MNRAS.487.2215A} {487, 2215}

\bibitem[\protect\citeauthoryear{{Arcavi} et~al.,}{{Arcavi} et~al.}{2011}]{Arcavi2011}
{Arcavi} I.,  et~al., 2011, \mn@doi [\apjl] {10.1088/2041-8205/742/2/L18}, \href {https://ui.adsabs.harvard.edu/abs/2011ApJ...742L..18A} {742, L18}

\bibitem[\protect\citeauthoryear{{Arnett}}{{Arnett}}{1982}]{Arnett1982}
{Arnett} W.~D.,  1982, \mn@doi [\apj] {10.1086/159681}, \href {https://ui.adsabs.harvard.edu/abs/1982ApJ...253..785A} {253, 785}

\bibitem[\protect\citeauthoryear{{Aryan} et~al.,}{{Aryan} et~al.}{2021}]{Aryan2021}
{Aryan} A.,  et~al., 2021, \mn@doi [\mnras] {10.1093/mnras/stab1379}, \href {https://ui.adsabs.harvard.edu/abs/2021MNRAS.505.2530A} {505, 2530}

\bibitem[\protect\citeauthoryear{{Barbarino} et~al.,}{{Barbarino} et~al.}{2015}]{Barbarino2015}
{Barbarino} C.,  et~al., 2015, \mn@doi [\mnras] {10.1093/mnras/stv106}, \href {https://ui.adsabs.harvard.edu/abs/2015MNRAS.448.2312B} {448, 2312}

\bibitem[\protect\citeauthoryear{{Barbary} et~al.,}{{Barbary} et~al.}{2016}]{Barbary2016}
{Barbary} K.,  et~al., 2016, Astrophysics Source Code Library, record ascl:1611.017, \href {https://ui.adsabs.harvard.edu/abs/2016ascl.soft11017B} {}

\bibitem[\protect\citeauthoryear{{Barbon}, {Benetti}, {Cappellaro}, {Patat}, {Turatto}  \& {Iijima}}{{Barbon} et~al.}{1995}]{Barbon1995}
{Barbon} R.,  {Benetti} S.,  {Cappellaro} E.,  {Patat} F.,  {Turatto} M.,   {Iijima} T.,  1995, \aaps, \href {https://ui.adsabs.harvard.edu/abs/1995A&AS..110..513B} {110, 513}

\bibitem[\protect\citeauthoryear{{Bayless} et~al.,}{{Bayless} et~al.}{2013}]{Bayless2013}
{Bayless} A.~J.,  et~al., 2013, \mn@doi [\apjl] {10.1088/2041-8205/764/1/L13}, \href {https://ui.adsabs.harvard.edu/abs/2013ApJ...764L..13B} {764, L13}

\bibitem[\protect\citeauthoryear{{Benetti} et~al.,}{{Benetti} et~al.}{2011}]{Benetti2011}
{Benetti} S.,  et~al., 2011, \mn@doi [\mnras] {10.1111/j.1365-2966.2010.17873.x}, \href {https://ui.adsabs.harvard.edu/abs/2011MNRAS.411.2726B} {411, 2726}

\bibitem[\protect\citeauthoryear{Bengyat \& Gal-Yam}{Bengyat \& Gal-Yam}{2022}]{Bengyat2022}
Bengyat O.,  Gal-Yam A.,  2022, The Astrophysical Journal, 930, 31

\bibitem[\protect\citeauthoryear{{Bessell}}{{Bessell}}{1990}]{Bessell1990}
{Bessell} M.~S.,  1990, \mn@doi [\pasp] {10.1086/132749}, \href {https://ui.adsabs.harvard.edu/abs/1990PASP..102.1181B} {102, 1181}

\bibitem[\protect\citeauthoryear{Bethe}{Bethe}{1990}]{Bethe1990}
Bethe H.~A.,  1990, \mn@doi [Rev. Mod. Phys.] {10.1103/RevModPhys.62.801}, 62, 801

\bibitem[\protect\citeauthoryear{{Bethe} \& {Wilson}}{{Bethe} \& {Wilson}}{1985}]{Bethe1985}
{Bethe} H.~A.,  {Wilson} J.~R.,  1985, \mn@doi [\apj] {10.1086/163343}, \href {https://ui.adsabs.harvard.edu/abs/1985ApJ...295...14B} {295, 14}

\bibitem[\protect\citeauthoryear{{Bianco} et~al.,}{{Bianco} et~al.}{2014}]{Bianco2014}
{Bianco} F.~B.,  et~al., 2014, \mn@doi [VizieR Online Data Catalog] {10.26093/cds/vizier.22130019}, \href {https://ui.adsabs.harvard.edu/abs/2014yCat..22130019B} {p. J/ApJS/213/19}

\bibitem[\protect\citeauthoryear{Bionta et~al.,}{Bionta et~al.}{1987}]{Bionta1987}
Bionta R.~M.,  et~al., 1987, \mn@doi [Phys. Rev. Lett.] {10.1103/PhysRevLett.58.1494}, 58, 1494

\bibitem[\protect\citeauthoryear{{Black}, {Milisavljevic}, {Margutti}, {Fesen}, {Patnaude}  \& {Parker}}{{Black} et~al.}{2017}]{Black2017}
{Black} C.~S.,  {Milisavljevic} D.,  {Margutti} R.,  {Fesen} R.~A.,  {Patnaude} D.,   {Parker} S.,  2017, \mn@doi [\apj] {10.3847/1538-4357/aa8999}, \href {https://ui.adsabs.harvard.edu/abs/2017ApJ...848....5B} {848, 5}

\bibitem[\protect\citeauthoryear{{Bose} et~al.,}{{Bose} et~al.}{2013}]{Bose2013}
{Bose} S.,  et~al., 2013, \mn@doi [\mnras] {10.1093/mnras/stt864}, \href {https://ui.adsabs.harvard.edu/abs/2013MNRAS.433.1871B} {433, 1871}

\bibitem[\protect\citeauthoryear{{Bose} et~al.,}{{Bose} et~al.}{2015a}]{Bose2015a}
{Bose} S.,  et~al., 2015a, \mn@doi [\mnras] {10.1093/mnras/stv759}, \href {https://ui.adsabs.harvard.edu/abs/2015MNRAS.450.2373B} {450, 2373}

\bibitem[\protect\citeauthoryear{{Bose} et~al.,}{{Bose} et~al.}{2015b}]{Bose2015b}
{Bose} S.,  et~al., 2015b, \mn@doi [\apj] {10.1088/0004-637X/806/2/160}, \href {https://ui.adsabs.harvard.edu/abs/2015ApJ...806..160B} {806, 160}

\bibitem[\protect\citeauthoryear{{Bose}, {Kumar}, {Misra}, {Matsumoto}, {Kumar}, {Singh}, {Fukushima}  \& {Kawabata}}{{Bose} et~al.}{2016}]{Bose2016}
{Bose} S.,  {Kumar} B.,  {Misra} K.,  {Matsumoto} K.,  {Kumar} B.,  {Singh} M.,  {Fukushima} D.,   {Kawabata} M.,  2016, \mn@doi [\mnras] {10.1093/mnras/stv2351}, \href {https://ui.adsabs.harvard.edu/abs/2016MNRAS.455.2712B} {455, 2712}

\bibitem[\protect\citeauthoryear{{Bostroem} et~al.,}{{Bostroem} et~al.}{2019}]{Bostroem2019}
{Bostroem} K.~A.,  et~al., 2019, \mn@doi [\mnras] {10.1093/mnras/stz570}, \href {https://ui.adsabs.harvard.edu/abs/2019MNRAS.485.5120B} {485, 5120}

\bibitem[\protect\citeauthoryear{{Branch} \& {Wheeler}}{{Branch} \& {Wheeler}}{2017}]{Branch2017}
{Branch} D.,  {Wheeler} J.~C.,  2017, {Supernova Explosions}.
Springer, \mn@doi{10.1007/978-3-662-55054-0}

\bibitem[\protect\citeauthoryear{{Braun}, {Bonaldi}, {Bourke}, {Keane}  \& {Wagg}}{{Braun} et~al.}{2019}]{Braun2019}
{Braun} R.,  {Bonaldi} A.,  {Bourke} T.,  {Keane} E.,   {Wagg} J.,  2019, \mn@doi [arXiv e-prints] {10.48550/arXiv.1912.12699}, \href {https://ui.adsabs.harvard.edu/abs/2019arXiv191212699B} {p. arXiv:1912.12699}

\bibitem[\protect\citeauthoryear{{Brown} et~al.,}{{Brown} et~al.}{2013}]{Brown2013}
{Brown} T.~M.,  et~al., 2013, \mn@doi [\pasp] {10.1086/673168}, \href {https://ui.adsabs.harvard.edu/abs/2013PASP..125.1031B} {125, 1031}

\bibitem[\protect\citeauthoryear{{Bufano} et~al.,}{{Bufano} et~al.}{2009}]{Bufano2009}
{Bufano} F.,  et~al., 2009, \mn@doi [\apj] {10.1088/0004-637X/700/2/1456}, \href {https://ui.adsabs.harvard.edu/abs/2009ApJ...700.1456B} {700, 1456}

\bibitem[\protect\citeauthoryear{{Bufano} et~al.,}{{Bufano} et~al.}{2014}]{Bufano2014}
{Bufano} F.,  et~al., 2014, \mn@doi [\mnras] {10.1093/mnras/stu065}, \href {https://ui.adsabs.harvard.edu/abs/2014MNRAS.439.1807B} {439, 1807}

\bibitem[\protect\citeauthoryear{{Bullivant} et~al.,}{{Bullivant} et~al.}{2018}]{Bullivant2018}
{Bullivant} C.,  et~al., 2018, \mn@doi [\mnras] {10.1093/mnras/sty045}, \href {https://ui.adsabs.harvard.edu/abs/2018MNRAS.476.1497B} {476, 1497}

\bibitem[\protect\citeauthoryear{{Buton} et~al.,}{{Buton} et~al.}{2013}]{Buton2013}
{Buton} C.,  et~al., 2013, \mn@doi [\aap] {10.1051/0004-6361/201219834}, \href {https://ui.adsabs.harvard.edu/abs/2013A&A...549A...8B} {549, A8}

\bibitem[\protect\citeauthoryear{{Cao} et~al.,}{{Cao} et~al.}{2013}]{Cao2013}
{Cao} Y.,  et~al., 2013, \mn@doi [\apjl] {10.1088/2041-8205/775/1/L7}, \href {https://ui.adsabs.harvard.edu/abs/2013ApJ...775L...7C} {775, L7}

\bibitem[\protect\citeauthoryear{{Cardelli}, {Clayton}  \& {Mathis}}{{Cardelli} et~al.}{1989}]{Cardelli1989}
{Cardelli} J.~A.,  {Clayton} G.~C.,   {Mathis} J.~S.,  1989, \mn@doi [\apj] {10.1086/167900}, \href {https://ui.adsabs.harvard.edu/abs/1989ApJ...345..245C} {345, 245}

\bibitem[\protect\citeauthoryear{{Catchpole} et~al.,}{{Catchpole} et~al.}{1987}]{Catchpole1987}
{Catchpole} R.~M.,  et~al., 1987, \mn@doi [\mnras] {10.1093/mnras/229.1.15P}, \href {https://ui.adsabs.harvard.edu/abs/1987MNRAS.229P..15C} {229, 15P}

\bibitem[\protect\citeauthoryear{{Catchpole} et~al.,}{{Catchpole} et~al.}{1988}]{Catchpole1988}
{Catchpole} R.~M.,  et~al., 1988, \mn@doi [\mnras] {10.1093/mnras/231.1.75P}, \href {https://ui.adsabs.harvard.edu/abs/1988MNRAS.231P..75C} {231, 75P}

\bibitem[\protect\citeauthoryear{{Catchpole} et~al.,}{{Catchpole} et~al.}{1989}]{Catchpole1989}
{Catchpole} R.~M.,  et~al., 1989, \mn@doi [\mnras] {10.1093/mnras/237.1.55P}, \href {https://ui.adsabs.harvard.edu/abs/1989MNRAS.237P..55C} {237, 55P}

\bibitem[\protect\citeauthoryear{{Charalampopoulos} et~al.,}{{Charalampopoulos} et~al.}{2022}]{Charalampopoulos2022}
{Charalampopoulos} P.,  et~al., 2022, \mn@doi [\aap] {10.1051/0004-6361/202142122}, \href {https://ui.adsabs.harvard.edu/abs/2022A&A...659A..34C} {659, A34}

\bibitem[\protect\citeauthoryear{{Chen} et~al.,}{{Chen} et~al.}{2014}]{Chen2014}
{Chen} J.,  et~al., 2014, \mn@doi [\apj] {10.1088/0004-637X/790/2/120}, \href {https://ui.adsabs.harvard.edu/abs/2014ApJ...790..120C} {790, 120}

\bibitem[\protect\citeauthoryear{{Chen} et~al.,}{{Chen} et~al.}{2020}]{Chen2020}
{Chen} P.,  et~al., 2020, \mn@doi [\apjl] {10.3847/2041-8213/ab62a4}, \href {https://ui.adsabs.harvard.edu/abs/2020ApJ...889L...6C} {889, L6}

\bibitem[\protect\citeauthoryear{{Chengalur}, {Salpeter}  \& {Terzian}}{{Chengalur} et~al.}{1993}]{Chengalur1993}
{Chengalur} J.~N.,  {Salpeter} E.~E.,   {Terzian} Y.,  1993, \mn@doi [\apj] {10.1086/173456}, \href {https://ui.adsabs.harvard.edu/abs/1993ApJ...419...30C} {419, 30}

\bibitem[\protect\citeauthoryear{{Childress} et~al.,}{{Childress} et~al.}{2016}]{Childress2016}
{Childress} M.~J.,  et~al., 2016, \mn@doi [\pasa] {10.1017/pasa.2016.47}, \href {https://ui.adsabs.harvard.edu/abs/2016PASA...33...55C} {33, e055}

\bibitem[\protect\citeauthoryear{{Chornock} et~al.,}{{Chornock} et~al.}{2013}]{Chornock2013}
{Chornock} R.,  et~al., 2013, \mn@doi [\apj] {10.1088/0004-637X/767/2/162}, \href {https://ui.adsabs.harvard.edu/abs/2013ApJ...767..162C} {767, 162}

\bibitem[\protect\citeauthoryear{{Clocchiatti}, {Wheeler}, {Brotherton}, {Cochran}, {Wills}, {Barker}  \& {Turatto}}{{Clocchiatti} et~al.}{1996}]{Clocchiatti1996}
{Clocchiatti} A.,  {Wheeler} J.~C.,  {Brotherton} M.~S.,  {Cochran} A.~L.,  {Wills} D.,  {Barker} E.~S.,   {Turatto} M.,  1996, \mn@doi [\apj] {10.1086/177165}, \href {https://ui.adsabs.harvard.edu/abs/1996ApJ...462..462C} {462, 462}

\bibitem[\protect\citeauthoryear{{Conley} et~al.,}{{Conley} et~al.}{2008}]{Conley2008}
{Conley} A.,  et~al., 2008, \mn@doi [\apj] {10.1086/588518}, \href {https://ui.adsabs.harvard.edu/abs/2008ApJ...681..482C} {681, 482}

\bibitem[\protect\citeauthoryear{{Couch}}{{Couch}}{2017}]{Couch2017}
{Couch} S.~M.,  2017, \mn@doi [Philosophical Transactions of the Royal Society of London Series A] {10.1098/rsta.2016.0271}, \href {https://ui.adsabs.harvard.edu/abs/2017RSPTA.37560271C} {375, 20160271}

\bibitem[\protect\citeauthoryear{{Dall'Ora} et~al.,}{{Dall'Ora} et~al.}{2014}]{Dall'Ora2014}
{Dall'Ora} M.,  et~al., 2014, \mn@doi [\apj] {10.1088/0004-637X/787/2/139}, \href {https://ui.adsabs.harvard.edu/abs/2014ApJ...787..139D} {787, 139}

\bibitem[\protect\citeauthoryear{{Davis} et~al.,}{{Davis} et~al.}{2021}]{Davis2021}
{Davis} S.,  et~al., 2021, \mn@doi [\apj] {10.3847/1538-4357/abdd36}, \href {https://ui.adsabs.harvard.edu/abs/2021ApJ...909..145D} {909, 145}

\bibitem[\protect\citeauthoryear{{Dhungana} et~al.,}{{Dhungana} et~al.}{2016}]{Dhungana2016}
{Dhungana} G.,  et~al., 2016, \mn@doi [\apj] {10.3847/0004-637X/822/1/6}, \href {https://ui.adsabs.harvard.edu/abs/2016ApJ...822....6D} {822, 6}

\bibitem[\protect\citeauthoryear{{Drout} et~al.,}{{Drout} et~al.}{2016}]{Drout2016}
{Drout} M.~R.,  et~al., 2016, \mn@doi [\apj] {10.3847/0004-637X/821/1/57}, \href {https://ui.adsabs.harvard.edu/abs/2016ApJ...821...57D} {821, 57}

\bibitem[\protect\citeauthoryear{{Eastman}, {Schmidt}  \& {Kirshner}}{{Eastman} et~al.}{1996}]{Eastman1996}
{Eastman} R.~G.,  {Schmidt} B.~P.,   {Kirshner} R.,  1996, \mn@doi [\apj] {10.1086/177563}, \href {https://ui.adsabs.harvard.edu/abs/1996ApJ...466..911E} {466, 911}

\bibitem[\protect\citeauthoryear{{Ebden}}{{Ebden}}{2015}]{Ebden2015}
{Ebden} M.,  2015, \mn@doi [arXiv e-prints] {10.48550/arXiv.1505.02965}, \href {https://ui.adsabs.harvard.edu/abs/2015arXiv150502965E} {p. arXiv:1505.02965}

\bibitem[\protect\citeauthoryear{{Elias-Rosa} et~al.,}{{Elias-Rosa} et~al.}{2010}]{Elias-Rosa2010}
{Elias-Rosa} N.,  et~al., 2010, \mn@doi [\apjl] {10.1088/2041-8205/714/2/L254}, \href {https://ui.adsabs.harvard.edu/abs/2010ApJ...714L.254E} {714, L254}

\bibitem[\protect\citeauthoryear{{Ergon} et~al.,}{{Ergon} et~al.}{2014}]{Ergon2014}
{Ergon} M.,  et~al., 2014, \mn@doi [\aap] {10.1051/0004-6361/201321850}, \href {https://ui.adsabs.harvard.edu/abs/2014A&A...562A..17E} {562, A17}

\bibitem[\protect\citeauthoryear{{Ergon} et~al.,}{{Ergon} et~al.}{2015}]{Ergon2015}
{Ergon} M.,  et~al., 2015, \mn@doi [\aap] {10.1051/0004-6361/201424592}, \href {https://ui.adsabs.harvard.edu/abs/2015A&A...580A.142E} {580, A142}

\bibitem[\protect\citeauthoryear{{Faran} et~al.,}{{Faran} et~al.}{2014a}]{Faran2014a}
{Faran} T.,  et~al., 2014a, \mn@doi [\mnras] {10.1093/mnras/stu955}, \href {https://ui.adsabs.harvard.edu/abs/2014MNRAS.442..844F} {442, 844}

\bibitem[\protect\citeauthoryear{{Faran} et~al.,}{{Faran} et~al.}{2014b}]{Faran2014b}
{Faran} T.,  et~al., 2014b, \mn@doi [\mnras] {10.1093/mnras/stu1760}, \href {https://ui.adsabs.harvard.edu/abs/2014MNRAS.445..554F} {445, 554}

\bibitem[\protect\citeauthoryear{{Filippenko}}{{Filippenko}}{1997}]{Filippenko1997}
{Filippenko} A.~V.,  1997, \mn@doi [\araa] {10.1146/annurev.astro.35.1.309}, \href {https://ui.adsabs.harvard.edu/abs/1997ARA&A..35..309F} {35, 309}

\bibitem[\protect\citeauthoryear{{Filippenko} et~al.,}{{Filippenko} et~al.}{1995}]{Filippenko1995}
{Filippenko} A.~V.,  et~al., 1995, \mn@doi [\apjl] {10.1086/309659}, \href {https://ui.adsabs.harvard.edu/abs/1995ApJ...450L..11F} {450, L11}

\bibitem[\protect\citeauthoryear{{Filippenko}, {Li}, {Treffers}  \& {Modjaz}}{{Filippenko} et~al.}{2001}]{Filippenko2001}
{Filippenko} A.~V.,  {Li} W.~D.,  {Treffers} R.~R.,   {Modjaz} M.,  2001, in {Paczynski} B.,  {Chen} W.-P.,   {Lemme} C.,  eds,  Astronomical Society of the Pacific Conference Series Vol. 246, IAU Colloq. 183: Small Telescope Astronomy on Global Scales. p.~121

\bibitem[\protect\citeauthoryear{Fiorillo, Heinlein, Janka, Raffelt, Vitagliano  \& Bollig}{Fiorillo et~al.}{2023}]{Fiorillo2023}
Fiorillo D.~F.,  Heinlein M.,  Janka H.-T.,  Raffelt G.,  Vitagliano E.,   Bollig R.,  2023, Physical Review D, 108, 083040

\bibitem[\protect\citeauthoryear{{Folatelli} et~al.,}{{Folatelli} et~al.}{2006}]{Folatelli2006}
{Folatelli} G.,  et~al., 2006, \mn@doi [\apj] {10.1086/500531}, \href {https://ui.adsabs.harvard.edu/abs/2006ApJ...641.1039F} {641, 1039}

\bibitem[\protect\citeauthoryear{{Foley} et~al.,}{{Foley} et~al.}{2003}]{Foley2003}
{Foley} R.~J.,  et~al., 2003, \mn@doi [\pasp] {10.1086/378242}, \href {https://ui.adsabs.harvard.edu/abs/2003PASP..115.1220F} {115, 1220}

\bibitem[\protect\citeauthoryear{{Fraser} et~al.,}{{Fraser} et~al.}{2013}]{Fraser2013}
{Fraser} M.,  et~al., 2013, \mn@doi [\mnras] {10.1093/mnras/stt813}, \href {https://ui.adsabs.harvard.edu/abs/2013MNRAS.433.1312F} {433, 1312}

\bibitem[\protect\citeauthoryear{{Fremling} et~al.,}{{Fremling} et~al.}{2014}]{Fremling2014}
{Fremling} C.,  et~al., 2014, \mn@doi [\aap] {10.1051/0004-6361/201423884}, \href {https://ui.adsabs.harvard.edu/abs/2014A&A...565A.114F} {565, A114}

\bibitem[\protect\citeauthoryear{{Fremling} et~al.,}{{Fremling} et~al.}{2016}]{Fremling2016}
{Fremling} C.,  et~al., 2016, \mn@doi [VizieR Online Data Catalog] {10.26093/cds/vizier.35930068}, \href {https://ui.adsabs.harvard.edu/abs/2016yCat..35930068F} {pp J/A+A/593/A68}

\bibitem[\protect\citeauthoryear{{Friedmann}}{{Friedmann}}{1922}]{Friedmann1922}
{Friedmann} A.,  1922, \mn@doi [Zeitschrift fur Physik] {10.1007/BF01332580}, \href {https://ui.adsabs.harvard.edu/abs/1922ZPhy...10..377F} {10, 377}

\bibitem[\protect\citeauthoryear{{Gal-Yam}, {Ofek}  \& {Shemmer}}{{Gal-Yam} et~al.}{2002}]{Gal-Yam2002}
{Gal-Yam} A.,  {Ofek} E.~O.,   {Shemmer} O.,  2002, \mn@doi [\mnras] {10.1046/j.1365-8711.2002.05535.x}, \href {https://ui.adsabs.harvard.edu/abs/2002MNRAS.332L..73G} {332, L73}

\bibitem[\protect\citeauthoryear{{Galama} et~al.,}{{Galama} et~al.}{1998}]{Galama1998}
{Galama} T.~J.,  et~al., 1998, \mn@doi [\nat] {10.1038/27150}, \href {https://ui.adsabs.harvard.edu/abs/1998Natur.395..670G} {395, 670}

\bibitem[\protect\citeauthoryear{{Galbany} et~al.,}{{Galbany} et~al.}{2016}]{Galbany2016}
{Galbany} L.,  et~al., 2016, \mn@doi [\aj] {10.3847/0004-6256/151/2/33}, \href {https://ui.adsabs.harvard.edu/abs/2016AJ....151...33G} {151, 33}

\bibitem[\protect\citeauthoryear{Gilmozzi \& Spyromilio}{Gilmozzi \& Spyromilio}{2007}]{Gilmozzi2007}
Gilmozzi R.,  Spyromilio J.,  2007, The Messenger, 127, 3

\bibitem[\protect\citeauthoryear{Glass}{Glass}{1973}]{Glass1973}
Glass I.,  1973, Monthly Notices of the Royal Astronomical Society, 164, 155

\bibitem[\protect\citeauthoryear{{Graham} et~al.,}{{Graham} et~al.}{2014}]{Graham2014}
{Graham} M.~L.,  et~al., 2014, \mn@doi [\apj] {10.1088/0004-637X/787/2/163}, \href {https://ui.adsabs.harvard.edu/abs/2014ApJ...787..163G} {787, 163}

\bibitem[\protect\citeauthoryear{{Guillochon}, {Parrent}, {Kelley}  \& {Margutti}}{{Guillochon} et~al.}{2017}]{Guillochon2017}
{Guillochon} J.,  {Parrent} J.,  {Kelley} L.~Z.,   {Margutti} R.,  2017, \mn@doi [\apj] {10.3847/1538-4357/835/1/64}, \href {https://ui.adsabs.harvard.edu/abs/2017ApJ...835...64G} {835, 64}

\bibitem[\protect\citeauthoryear{Gunn et~al.,}{Gunn et~al.}{1998}]{Gunn1998}
Gunn J.~E.,  et~al., 1998, The Astronomical Journal, 116, 3040

\bibitem[\protect\citeauthoryear{{Hamuy} et~al.,}{{Hamuy} et~al.}{2001}]{Hamuy2001}
{Hamuy} M.,  et~al., 2001, \mn@doi [\apj] {10.1086/322450}, \href {https://ui.adsabs.harvard.edu/abs/2001ApJ...558..615H} {558, 615}

\bibitem[\protect\citeauthoryear{Harris et~al.,}{Harris et~al.}{2020}]{Harris2020}
Harris C.~R.,  et~al., 2020, \mn@doi [Nature] {10.1038/s41586-020-2649-2}, 585, 357

\bibitem[\protect\citeauthoryear{{Harutyunyan} et~al.,}{{Harutyunyan} et~al.}{2008}]{Harutyunyan2008}
{Harutyunyan} A.~H.,  et~al., 2008, \mn@doi [\aap] {10.1051/0004-6361:20078859}, \href {https://ui.adsabs.harvard.edu/abs/2008A&A...488..383H} {488, 383}

\bibitem[\protect\citeauthoryear{{Hicken} et~al.,}{{Hicken} et~al.}{2017}]{Hicken2017}
{Hicken} M.,  et~al., 2017, \mn@doi [\apjs] {10.3847/1538-4365/aa8ef4}, \href {https://ui.adsabs.harvard.edu/abs/2017ApJS..233....6H} {233, 6}

\bibitem[\protect\citeauthoryear{Hirata et~al.,}{Hirata et~al.}{1987}]{Hirata1987}
Hirata K.,  et~al., 1987, \mn@doi [Phys. Rev. Lett.] {10.1103/PhysRevLett.58.1490}, 58, 1490

\bibitem[\protect\citeauthoryear{{Hosseinzadeh} et~al.,}{{Hosseinzadeh} et~al.}{2018}]{Hosseinzadeh2018}
{Hosseinzadeh} G.,  et~al., 2018, \mn@doi [\apj] {10.3847/1538-4357/aac5f6}, \href {https://ui.adsabs.harvard.edu/abs/2018ApJ...861...63H} {861, 63}

\bibitem[\protect\citeauthoryear{{Hosseinzadeh} et~al.,}{{Hosseinzadeh} et~al.}{2022}]{Hosseinzadeh2022}
{Hosseinzadeh} G.,  et~al., 2022, \mn@doi [\apj] {10.3847/1538-4357/ac75f0}, \href {https://ui.adsabs.harvard.edu/abs/2022ApJ...935...31H} {935, 31}

\bibitem[\protect\citeauthoryear{{Hsiao}, {Conley}, {Howell}, {Sullivan}, {Pritchet}, {Carlberg}, {Nugent}  \& {Phillips}}{{Hsiao} et~al.}{2007}]{Hsiao2007}
{Hsiao} E.~Y.,  {Conley} A.,  {Howell} D.~A.,  {Sullivan} M.,  {Pritchet} C.~J.,  {Carlberg} R.~G.,  {Nugent} P.~E.,   {Phillips} M.~M.,  2007, \mn@doi [\apj] {10.1086/518232}, \href {https://ui.adsabs.harvard.edu/abs/2007ApJ...663.1187H} {663, 1187}

\bibitem[\protect\citeauthoryear{{Huang} et~al.,}{{Huang} et~al.}{2018}]{Huang2018}
{Huang} F.,  et~al., 2018, \mn@doi [\mnras] {10.1093/mnras/sty066}, \href {https://ui.adsabs.harvard.edu/abs/2018MNRAS.475.3959H} {475, 3959}

\bibitem[\protect\citeauthoryear{{Hubble}}{{Hubble}}{1929}]{Hubble1929}
{Hubble} E.,  1929, \mn@doi [Proceedings of the National Academy of Science] {10.1073/pnas.15.3.168}, \href {https://ui.adsabs.harvard.edu/abs/1929PNAS...15..168H} {15, 168}

\bibitem[\protect\citeauthoryear{{Humphreys}, {Davidson}, {Jones}, {Pogge}, {Grammer}, {Prieto}  \& {Pritchard}}{{Humphreys} et~al.}{2012}]{Humphreys2012}
{Humphreys} R.~M.,  {Davidson} K.,  {Jones} T.~J.,  {Pogge} R.~W.,  {Grammer} S.~H.,  {Prieto} J.~L.,   {Pritchard} T.~A.,  2012, \mn@doi [\apj] {10.1088/0004-637X/760/1/93}, \href {https://ui.adsabs.harvard.edu/abs/2012ApJ...760...93H} {760, 93}

\bibitem[\protect\citeauthoryear{{Inserra} et~al.,}{{Inserra} et~al.}{2011}]{Inserra2011}
{Inserra} C.,  et~al., 2011, \mn@doi [\mnras] {10.1111/j.1365-2966.2011.19128.x}, \href {https://ui.adsabs.harvard.edu/abs/2011MNRAS.417..261I} {417, 261}

\bibitem[\protect\citeauthoryear{{Inserra} et~al.,}{{Inserra} et~al.}{2012}]{Inserra2012}
{Inserra} C.,  et~al., 2012, \mn@doi [\mnras] {10.1111/j.1365-2966.2012.20685.x}, \href {https://ui.adsabs.harvard.edu/abs/2012MNRAS.422.1122I} {422, 1122}

\bibitem[\protect\citeauthoryear{{Inserra} et~al.,}{{Inserra} et~al.}{2013}]{Inserra2013}
{Inserra} C.,  et~al., 2013, \mn@doi [\aap] {10.1051/0004-6361/201220496}, \href {https://ui.adsabs.harvard.edu/abs/2013A&A...555A.142I} {555, A142}

\bibitem[\protect\citeauthoryear{{Inserra}, {Prajs}, {Gutierrez}, {Angus}, {Smith}  \& {Sullivan}}{{Inserra} et~al.}{2018}]{Inserra2018}
{Inserra} C.,  {Prajs} S.,  {Gutierrez} C.~P.,  {Angus} C.,  {Smith} M.,   {Sullivan} M.,  2018, \mn@doi [\apj] {10.3847/1538-4357/aaaaaa}, \href {https://ui.adsabs.harvard.edu/abs/2018ApJ...854..175I} {854, 175}

\bibitem[\protect\citeauthoryear{{Irani} et~al.,}{{Irani} et~al.}{2022}]{Irani2022}
{Irani} I.,  et~al., 2022, \mn@doi [\apj] {10.3847/1538-4357/ac4709}, \href {https://ui.adsabs.harvard.edu/abs/2022ApJ...927...10I} {927, 10}

\bibitem[\protect\citeauthoryear{Irani et~al.,}{Irani et~al.}{2024}]{Irani2024}
Irani I.,  et~al., 2024, The Early Ultraviolet Light-Curves of Type II Supernovae and the Radii of Their Progenitor Stars (\mn@eprint {arXiv} {2310.16885})

\bibitem[\protect\citeauthoryear{{Ivezi{\'c}} et~al.,}{{Ivezi{\'c}} et~al.}{2019}]{Ivezić2019}
{Ivezi{\'c}} {\v{Z}}.,  et~al., 2019, \mn@doi [\apj] {10.3847/1538-4357/ab042c}, \href {https://ui.adsabs.harvard.edu/abs/2019ApJ...873..111I} {873, 111}

\bibitem[\protect\citeauthoryear{{Janka}}{{Janka}}{2017}]{Janka2017}
{Janka} H.-T.,  2017, \mn@doi [Handbook of Supernovae] {10.1007/978-3-319-21846-5_109}, \href {https://ui.adsabs.harvard.edu/abs/2017hsn..book.1095J} {p.~1095}

\bibitem[\protect\citeauthoryear{Johnson}{Johnson}{1962}]{Johnson1962}
Johnson H.~L.,  1962, Astrophysical Journal, vol. 135, p. 69, 135, 69

\bibitem[\protect\citeauthoryear{{Kangas} et~al.,}{{Kangas} et~al.}{2016}]{Kangas2016}
{Kangas} T.,  et~al., 2016, \mn@doi [\mnras] {10.1093/mnras/stv2567}, \href {https://ui.adsabs.harvard.edu/abs/2016MNRAS.456..323K} {456, 323}

\bibitem[\protect\citeauthoryear{{Kankare} et~al.,}{{Kankare} et~al.}{2021}]{Kankare2021}
{Kankare} E.,  et~al., 2021, \mn@doi [\aap] {10.1051/0004-6361/202039240}, \href {https://ui.adsabs.harvard.edu/abs/2021A&A...649A.134K} {649, A134}

\bibitem[\protect\citeauthoryear{{Karamehmetoglu} et~al.,}{{Karamehmetoglu} et~al.}{2023}]{Karamehmetoglu2023}
{Karamehmetoglu} E.,  et~al., 2023, \mn@doi [\aap] {10.1051/0004-6361/202245231}, \href {https://ui.adsabs.harvard.edu/abs/2023A&A...678A..87K} {678, A87}

\bibitem[\protect\citeauthoryear{{Kessler} et~al.,}{{Kessler} et~al.}{2009}]{Kessler2009}
{Kessler} R.,  et~al., 2009, \mn@doi [\pasp] {10.1086/605984}, \href {https://ui.adsabs.harvard.edu/abs/2009PASP..121.1028K} {121, 1028}

\bibitem[\protect\citeauthoryear{{Kessler} et~al.,}{{Kessler} et~al.}{2010}]{Kessler2010}
{Kessler} R.,  et~al., 2010, \mn@doi [\pasp] {10.1086/657607}, \href {https://ui.adsabs.harvard.edu/abs/2010PASP..122.1415K} {122, 1415}

\bibitem[\protect\citeauthoryear{{Khazov} et~al.,}{{Khazov} et~al.}{2016}]{Khazov2016}
{Khazov} D.,  et~al., 2016, \mn@doi [\apj] {10.3847/0004-637X/818/1/3}, \href {https://ui.adsabs.harvard.edu/abs/2016ApJ...818....3K} {818, 3}

\bibitem[\protect\citeauthoryear{{Kim} et~al.,}{{Kim} et~al.}{2013}]{Kim2013}
{Kim} A.~G.,  et~al., 2013, \mn@doi [\apj] {10.1088/0004-637X/766/2/84}, \href {https://ui.adsabs.harvard.edu/abs/2013ApJ...766...84K} {766, 84}

\bibitem[\protect\citeauthoryear{{Kumar} et~al.,}{{Kumar} et~al.}{2013}]{Kumar2013}
{Kumar} B.,  et~al., 2013, \mn@doi [\mnras] {10.1093/mnras/stt162}, \href {https://ui.adsabs.harvard.edu/abs/2013MNRAS.431..308K} {431, 308}

\bibitem[\protect\citeauthoryear{{Kuncarayakti} et~al.,}{{Kuncarayakti} et~al.}{2018}]{Kuncarayakti2018}
{Kuncarayakti} H.,  et~al., 2018, \mn@doi [\apjl] {10.3847/2041-8213/aaaa1a}, \href {https://ui.adsabs.harvard.edu/abs/2018ApJ...854L..14K} {854, L14}

\bibitem[\protect\citeauthoryear{{Leonard} et~al.,}{{Leonard} et~al.}{2002}]{Leonard2002}
{Leonard} D.~C.,  et~al., 2002, \mn@doi [\pasp] {10.1086/324785}, \href {https://ui.adsabs.harvard.edu/abs/2002PASP..114...35L} {114, 35}

\bibitem[\protect\citeauthoryear{{Li}, {Chornock}, {Leaman}, {Filippenko}, {Poznanski}, {Wang}, {Ganeshalingam}  \& {Mannucci}}{{Li} et~al.}{2011}]{Li2011}
{Li} W.,  {Chornock} R.,  {Leaman} J.,  {Filippenko} A.~V.,  {Poznanski} D.,  {Wang} X.,  {Ganeshalingam} M.,   {Mannucci} F.,  2011, \mn@doi [\mnras] {10.1111/j.1365-2966.2011.18162.x}, \href {https://ui.adsabs.harvard.edu/abs/2011MNRAS.412.1473L} {412, 1473}

\bibitem[\protect\citeauthoryear{{Lochner}, {McEwen}, {Peiris}, {Lahav}  \& {Winter}}{{Lochner} et~al.}{2016}]{Lochner2016}
{Lochner} M.,  {McEwen} J.~D.,  {Peiris} H.~V.,  {Lahav} O.,   {Winter} M.~K.,  2016, \mn@doi [\apjs] {10.3847/0067-0049/225/2/31}, \href {https://ui.adsabs.harvard.edu/abs/2016ApJS..225...31L} {225, 31}

\bibitem[\protect\citeauthoryear{{Lusk} \& {Baron}}{{Lusk} \& {Baron}}{2017}]{Lusk2017}
{Lusk} J.~A.,  {Baron} E.,  2017, \mn@doi [\pasp] {10.1088/1538-3873/aa5e49}, \href {https://ui.adsabs.harvard.edu/abs/2017PASP..129d4202L} {129, 044202}

\bibitem[\protect\citeauthoryear{{Maguire} et~al.,}{{Maguire} et~al.}{2010}]{Maguire2010}
{Maguire} K.,  et~al., 2010, \mn@doi [\mnras] {10.1111/j.1365-2966.2010.16332.x}, \href {https://ui.adsabs.harvard.edu/abs/2010MNRAS.404..981M} {404, 981}

\bibitem[\protect\citeauthoryear{{Margutti} et~al.,}{{Margutti} et~al.}{2014}]{Margutti2014}
{Margutti} R.,  et~al., 2014, \mn@doi [\apj] {10.1088/0004-637X/780/1/21}, \href {https://ui.adsabs.harvard.edu/abs/2014ApJ...780...21M} {780, 21}

\bibitem[\protect\citeauthoryear{{Matheson} et~al.,}{{Matheson} et~al.}{2000a}]{Matheson2000a}
{Matheson} T.,  et~al., 2000a, \mn@doi [\aj] {10.1086/301518}, \href {https://ui.adsabs.harvard.edu/abs/2000AJ....120.1487M} {120, 1487}

\bibitem[\protect\citeauthoryear{{Matheson}, {Filippenko}, {Ho}, {Barth}  \& {Leonard}}{{Matheson} et~al.}{2000b}]{Matheson2000b}
{Matheson} T.,  {Filippenko} A.~V.,  {Ho} L.~C.,  {Barth} A.~J.,   {Leonard} D.~C.,  2000b, \mn@doi [\aj] {10.1086/301519}, \href {https://ui.adsabs.harvard.edu/abs/2000AJ....120.1499M} {120, 1499}

\bibitem[\protect\citeauthoryear{{Mauerhan} et~al.,}{{Mauerhan} et~al.}{2013a}]{Mauerhan2013a}
{Mauerhan} J.~C.,  et~al., 2013a, \mn@doi [\mnras] {10.1093/mnras/stt009}, \href {https://ui.adsabs.harvard.edu/abs/2013MNRAS.430.1801M} {430, 1801}

\bibitem[\protect\citeauthoryear{{Mauerhan} et~al.,}{{Mauerhan} et~al.}{2013b}]{Mauerhan2013b}
{Mauerhan} J.~C.,  et~al., 2013b, \mn@doi [\mnras] {10.1093/mnras/stt360}, \href {https://ui.adsabs.harvard.edu/abs/2013MNRAS.431.2599M} {431, 2599}

\bibitem[\protect\citeauthoryear{{Maund} et~al.,}{{Maund} et~al.}{2013}]{Maund2013}
{Maund} J.~R.,  et~al., 2013, \mn@doi [\mnras] {10.1093/mnrasl/slt017}, \href {https://ui.adsabs.harvard.edu/abs/2013MNRAS.431L.102M} {431, L102}

\bibitem[\protect\citeauthoryear{{Mazzali} et~al.,}{{Mazzali} et~al.}{2008}]{Mazzali2008}
{Mazzali} P.~A.,  et~al., 2008, \mn@doi [Science] {10.1126/science.1158088}, \href {https://ui.adsabs.harvard.edu/abs/2008Sci...321.1185M} {321, 1185}

\bibitem[\protect\citeauthoryear{{McBrien} et~al.,}{{McBrien} et~al.}{2019}]{McBrien2019}
{McBrien} O.~R.,  et~al., 2019, \mn@doi [\apjl] {10.3847/2041-8213/ab4dae}, \href {https://ui.adsabs.harvard.edu/abs/2019ApJ...885L..23M} {885, L23}

\bibitem[\protect\citeauthoryear{{Meza} et~al.,}{{Meza} et~al.}{2019}]{Meza2019}
{Meza} N.,  et~al., 2019, \mn@doi [\aap] {10.1051/0004-6361/201834972}, \href {https://ui.adsabs.harvard.edu/abs/2019A&A...629A..57M} {629, A57}

\bibitem[\protect\citeauthoryear{{Milisavljevic}, {Fesen}, {Nordsieck}, {Pickering}, {Gulbis}  \& {O'Donoghue}}{{Milisavljevic} et~al.}{2011}]{Milisavljevic2011}
{Milisavljevic} D.,  {Fesen} R.,  {Nordsieck} K.,  {Pickering} T.,  {Gulbis} A.,   {O'Donoghue} D.,  2011, Central Bureau Electronic Telegrams, \href {https://ui.adsabs.harvard.edu/abs/2011CBET.2777....3M} {2777, 3}

\bibitem[\protect\citeauthoryear{{Milisavljevic} et~al.,}{{Milisavljevic} et~al.}{2013a}]{Milisavljevic2013a}
{Milisavljevic} D.,  et~al., 2013a, \mn@doi [\apj] {10.1088/0004-637X/767/1/71}, \href {https://ui.adsabs.harvard.edu/abs/2013ApJ...767...71M} {767, 71}

\bibitem[\protect\citeauthoryear{{Milisavljevic} et~al.,}{{Milisavljevic} et~al.}{2013b}]{Milisavljevic2013b}
{Milisavljevic} D.,  et~al., 2013b, \mn@doi [\apjl] {10.1088/2041-8205/770/2/L38}, \href {https://ui.adsabs.harvard.edu/abs/2013ApJ...770L..38M} {770, L38}

\bibitem[\protect\citeauthoryear{{Milisavljevic} et~al.,}{{Milisavljevic} et~al.}{2015}]{Milisavljevic2015}
{Milisavljevic} D.,  et~al., 2015, \mn@doi [\apj] {10.1088/0004-637X/799/1/51}, \href {https://ui.adsabs.harvard.edu/abs/2015ApJ...799...51M} {799, 51}

\bibitem[\protect\citeauthoryear{{Modjaz} et~al.,}{{Modjaz} et~al.}{2006}]{Modjaz2006}
{Modjaz} M.,  et~al., 2006, \mn@doi [\apjl] {10.1086/505906}, \href {https://ui.adsabs.harvard.edu/abs/2006ApJ...645L..21M} {645, L21}

\bibitem[\protect\citeauthoryear{{Modjaz}, {Kirshner}, {Blondin}, {Challis}  \& {Matheson}}{{Modjaz} et~al.}{2008}]{Modjaz2008}
{Modjaz} M.,  {Kirshner} R.~P.,  {Blondin} S.,  {Challis} P.,   {Matheson} T.,  2008, \mn@doi [\apjl] {10.1086/593135}, \href {https://ui.adsabs.harvard.edu/abs/2008ApJ...687L...9M} {687, L9}

\bibitem[\protect\citeauthoryear{{Modjaz} et~al.,}{{Modjaz} et~al.}{2009}]{Modjaz2009}
{Modjaz} M.,  et~al., 2009, \mn@doi [\apj] {10.1088/0004-637X/702/1/226}, \href {https://ui.adsabs.harvard.edu/abs/2009ApJ...702..226M} {702, 226}

\bibitem[\protect\citeauthoryear{{Modjaz} et~al.,}{{Modjaz} et~al.}{2014}]{Modjaz2014}
{Modjaz} M.,  et~al., 2014, \mn@doi [\aj] {10.1088/0004-6256/147/5/99}, \href {https://ui.adsabs.harvard.edu/abs/2014AJ....147...99M} {147, 99}

\bibitem[\protect\citeauthoryear{M{\"o}ller \& de Boissi{\`e}re}{M{\"o}ller \& de~Boissi{\`e}re}{2020}]{Moller2020}
M{\"o}ller A.,  de Boissi{\`e}re T.,  2020, Monthly Notices of the Royal Astronomical Society, 491, 4277

\bibitem[\protect\citeauthoryear{{Morales-Garoffolo} et~al.,}{{Morales-Garoffolo} et~al.}{2014}]{Morales-Garoffolo2014}
{Morales-Garoffolo} A.,  et~al., 2014, \mn@doi [\mnras] {10.1093/mnras/stu1837}, \href {https://ui.adsabs.harvard.edu/abs/2014MNRAS.445.1647M} {445, 1647}

\bibitem[\protect\citeauthoryear{{Morales-Garoffolo} et~al.,}{{Morales-Garoffolo} et~al.}{2015}]{Morales-Garoffolo2015}
{Morales-Garoffolo} A.,  et~al., 2015, \mn@doi [\mnras] {10.1093/mnras/stv1972}, \href {https://ui.adsabs.harvard.edu/abs/2015MNRAS.454...95M} {454, 95}

\bibitem[\protect\citeauthoryear{{Munari}, {Henden}, {Belligoli}, {Castellani}, {Cherini}, {Righetti}  \& {Vagnozzi}}{{Munari} et~al.}{2013}]{Munari2013}
{Munari} U.,  {Henden} A.,  {Belligoli} R.,  {Castellani} F.,  {Cherini} G.,  {Righetti} G.~L.,   {Vagnozzi} A.,  2013, \mn@doi [\na] {10.1016/j.newast.2012.09.003}, \href {https://ui.adsabs.harvard.edu/abs/2013NewA...20...30M} {20, 30}

\bibitem[\protect\citeauthoryear{{Muthukrishna}, {Narayan}, {Mandel}, {Biswas}  \& {Hlo{\v{z}}ek}}{{Muthukrishna} et~al.}{2019}]{Muthukrishna2019}
{Muthukrishna} D.,  {Narayan} G.,  {Mandel} K.~S.,  {Biswas} R.,   {Hlo{\v{z}}ek} R.,  2019, \mn@doi [\pasp] {10.1088/1538-3873/ab1609}, \href {https://ui.adsabs.harvard.edu/abs/2019PASP..131k8002M} {131, 118002}

\bibitem[\protect\citeauthoryear{{Nakamura}, {Horiuchi}, {Tanaka}, {Hayama}, {Takiwaki}  \& {Kotake}}{{Nakamura} et~al.}{2016}]{Nakamura2016}
{Nakamura} K.,  {Horiuchi} S.,  {Tanaka} M.,  {Hayama} K.,  {Takiwaki} T.,   {Kotake} K.,  2016, \mn@doi [\mnras] {10.1093/mnras/stw1453}, \href {https://ui.adsabs.harvard.edu/abs/2016MNRAS.461.3296N} {461, 3296}

\bibitem[\protect\citeauthoryear{{Nakaoka} et~al.,}{{Nakaoka} et~al.}{2018}]{Nakaoka2018}
{Nakaoka} T.,  et~al., 2018, \mn@doi [\apj] {10.3847/1538-4357/aabee7}, \href {https://ui.adsabs.harvard.edu/abs/2018ApJ...859...78N} {859, 78}

\bibitem[\protect\citeauthoryear{Nicholl}{Nicholl}{2018}]{Nicholl2018}
Nicholl M.,  2018, \mn@doi [Research Notes of the AAS] {10.3847/2515-5172/aaf799}, 2, 230

\bibitem[\protect\citeauthoryear{{Nomoto}, {Yamaoka}, {Pols}, {van den Heuvel}, {Iwamoto}, {Kumagai}  \& {Shigeyama}}{{Nomoto} et~al.}{1994}]{Nomoto1994}
{Nomoto} K.,  {Yamaoka} H.,  {Pols} O.~R.,  {van den Heuvel} E.~P.~J.,  {Iwamoto} K.,  {Kumagai} S.,   {Shigeyama} T.,  1994, \mn@doi [\nat] {10.1038/371227a0}, \href {https://ui.adsabs.harvard.edu/abs/1994Natur.371..227N} {371, 227}

\bibitem[\protect\citeauthoryear{{Ochsenbein}, {Bauer}  \& {Marcout}}{{Ochsenbein} et~al.}{2000}]{Ochsenbein2000}
{Ochsenbein} F.,  {Bauer} P.,   {Marcout} J.,  2000, \mn@doi [\aaps] {10.1051/aas:2000169}, \href {https://ui.adsabs.harvard.edu/abs/2000A&AS..143...23O} {143, 23}

\bibitem[\protect\citeauthoryear{{Pandey}, {Anupama}, {Sagar}, {Bhattacharya}, {Sahu}  \& {Pandey}}{{Pandey} et~al.}{2003}]{Pandey2003}
{Pandey} S.~B.,  {Anupama} G.~C.,  {Sagar} R.,  {Bhattacharya} D.,  {Sahu} D.~K.,   {Pandey} J.~C.,  2003, \mn@doi [\mnras] {10.1046/j.1365-8711.2003.06148.x}, \href {https://ui.adsabs.harvard.edu/abs/2003MNRAS.340..375P} {340, 375}

\bibitem[\protect\citeauthoryear{{Pastorello} et~al.,}{{Pastorello} et~al.}{2006}]{Pastorello2006}
{Pastorello} A.,  et~al., 2006, \mn@doi [\mnras] {10.1111/j.1365-2966.2006.10587.x}, \href {https://ui.adsabs.harvard.edu/abs/2006MNRAS.370.1752P} {370, 1752}

\bibitem[\protect\citeauthoryear{{Pastorello} et~al.,}{{Pastorello} et~al.}{2008}]{Pastorello2008}
{Pastorello} A.,  et~al., 2008, \mn@doi [\mnras] {10.1111/j.1365-2966.2008.13618.x}, \href {https://ui.adsabs.harvard.edu/abs/2008MNRAS.389..955P} {389, 955}

\bibitem[\protect\citeauthoryear{{Pastorello} et~al.,}{{Pastorello} et~al.}{2009}]{Pastorello2009}
{Pastorello} A.,  et~al., 2009, \mn@doi [\mnras] {10.1111/j.1365-2966.2009.14505.x}, \href {https://ui.adsabs.harvard.edu/abs/2009MNRAS.394.2266P} {394, 2266}

\bibitem[\protect\citeauthoryear{{Pastorello} et~al.,}{{Pastorello} et~al.}{2015a}]{Pastorello2015}
{Pastorello} A.,  et~al., 2015a, \mn@doi [\mnras] {10.1093/mnras/stu2745}, \href {https://ui.adsabs.harvard.edu/abs/2015MNRAS.449.1921P} {449, 1921}

\bibitem[\protect\citeauthoryear{{Pastorello} et~al.,}{{Pastorello} et~al.}{2015b}]{Pastorello2015b}
{Pastorello} A.,  et~al., 2015b, \mn@doi [\mnras] {10.1093/mnras/stv1812}, \href {https://ui.adsabs.harvard.edu/abs/2015MNRAS.453.3649P} {453, 3649}

\bibitem[\protect\citeauthoryear{{Pastorello} et~al.,}{{Pastorello} et~al.}{2018}]{Pastorello2018}
{Pastorello} A.,  et~al., 2018, \mn@doi [\mnras] {10.1093/mnras/stx2668}, \href {https://ui.adsabs.harvard.edu/abs/2018MNRAS.474..197P} {474, 197}

\bibitem[\protect\citeauthoryear{{Patat} et~al.,}{{Patat} et~al.}{2001}]{Patat2001}
{Patat} F.,  et~al., 2001, \mn@doi [\apj] {10.1086/321526}, \href {https://ui.adsabs.harvard.edu/abs/2001ApJ...555..900P} {555, 900}

\bibitem[\protect\citeauthoryear{Pearson}{Pearson}{1900}]{Pearson1900}
Pearson K.,  1900, The London, Edinburgh, and Dublin Philosophical Magazine and Journal of Science, 50, 157

\bibitem[\protect\citeauthoryear{Pedregosa et~al.,}{Pedregosa et~al.}{2011}]{Pedregosa2011}
Pedregosa F.,  et~al., 2011, Journal of Machine Learning Research, 12, 2825

\bibitem[\protect\citeauthoryear{{Phillips}, {Lira}, {Suntzeff}, {Schommer}, {Hamuy}  \& {Maza}}{{Phillips} et~al.}{1999}]{Phillips1999}
{Phillips} M.~M.,  {Lira} P.,  {Suntzeff} N.~B.,  {Schommer} R.~A.,  {Hamuy} M.,   {Maza} J.,  1999, \mn@doi [\aj] {10.1086/301032}, \href {https://ui.adsabs.harvard.edu/abs/1999AJ....118.1766P} {118, 1766}

\bibitem[\protect\citeauthoryear{{Pian} et~al.,}{{Pian} et~al.}{2006}]{Pian2006}
{Pian} E.,  et~al., 2006, \mn@doi [\nat] {10.1038/nature05082}, \href {https://ui.adsabs.harvard.edu/abs/2006Natur.442.1011P} {442, 1011}

\bibitem[\protect\citeauthoryear{{Pian} et~al.,}{{Pian} et~al.}{2020}]{Pian2020}
{Pian} E.,  et~al., 2020, \mn@doi [\mnras] {10.1093/mnras/staa2191}, \href {https://ui.adsabs.harvard.edu/abs/2020MNRAS.497.3542P} {497, 3542}

\bibitem[\protect\citeauthoryear{{Pignata} et~al.,}{{Pignata} et~al.}{2009}]{Pignata2009}
{Pignata} G.,  et~al., 2009, Central Bureau Electronic Telegrams, \href {https://ui.adsabs.harvard.edu/abs/2009CBET.1731....1P} {1731, 1}

\bibitem[\protect\citeauthoryear{{Pignata} et~al.,}{{Pignata} et~al.}{2011}]{Pignata2011}
{Pignata} G.,  et~al., 2011, \mn@doi [\apj] {10.1088/0004-637X/728/1/14}, \href {https://ui.adsabs.harvard.edu/abs/2011ApJ...728...14P} {728, 14}

\bibitem[\protect\citeauthoryear{Poole et~al.,}{Poole et~al.}{2008}]{Poole2008}
Poole T.,  et~al., 2008, Monthly Notices of the Royal Astronomical Society, 383, 627

\bibitem[\protect\citeauthoryear{{Prentice} et~al.,}{{Prentice} et~al.}{2018a}]{Prentice2018a}
{Prentice} S.~J.,  et~al., 2018a, \mn@doi [\mnras] {10.1093/mnras/sty1223}, \href {https://ui.adsabs.harvard.edu/abs/2018MNRAS.478.4162P} {478, 4162}

\bibitem[\protect\citeauthoryear{{Prentice} et~al.,}{{Prentice} et~al.}{2018b}]{Prentice2018b}
{Prentice} S.~J.,  et~al., 2018b, \mn@doi [\apjl] {10.3847/2041-8213/aadd90}, \href {https://ui.adsabs.harvard.edu/abs/2018ApJ...865L...3P} {865, L3}

\bibitem[\protect\citeauthoryear{{Prentice} et~al.,}{{Prentice} et~al.}{2019}]{Prentice2019}
{Prentice} S.~J.,  et~al., 2019, \mn@doi [\mnras] {10.1093/mnras/sty3399}, \href {https://ui.adsabs.harvard.edu/abs/2019MNRAS.485.1559P} {485, 1559}

\bibitem[\protect\citeauthoryear{{Prentice} et~al.,}{{Prentice} et~al.}{2020}]{Prentice2020}
{Prentice} S.~J.,  et~al., 2020, \mn@doi [\aap] {10.1051/0004-6361/201936515}, \href {https://ui.adsabs.harvard.edu/abs/2020A&A...635A.186P} {635, A186}

\bibitem[\protect\citeauthoryear{{Pun} et~al.,}{{Pun} et~al.}{1995}]{Pun1995}
{Pun} C. S.~J.,  et~al., 1995, \mn@doi [\apjs] {10.1086/192185}, \href {https://ui.adsabs.harvard.edu/abs/1995ApJS...99..223P} {99, 223}

\bibitem[\protect\citeauthoryear{{Qu}, {Sako}, {M{\"o}ller}  \& {Doux}}{{Qu} et~al.}{2021}]{Qu2021}
{Qu} H.,  {Sako} M.,  {M{\"o}ller} A.,   {Doux} C.,  2021, \mn@doi [\aj] {10.3847/1538-3881/ac0824}, \href {https://ui.adsabs.harvard.edu/abs/2021AJ....162...67Q} {162, 67}

\bibitem[\protect\citeauthoryear{{Quimby} et~al.,}{{Quimby} et~al.}{2018}]{Quimby2018}
{Quimby} R.~M.,  et~al., 2018, \mn@doi [\apj] {10.3847/1538-4357/aaac2f}, \href {https://ui.adsabs.harvard.edu/abs/2018ApJ...855....2Q} {855, 2}

\bibitem[\protect\citeauthoryear{{Rabinak} \& {Waxman}}{{Rabinak} \& {Waxman}}{2011}]{Rabinak2011}
{Rabinak} I.,  {Waxman} E.,  2011, \mn@doi [\apj] {10.1088/0004-637X/728/1/63}, \href {https://ui.adsabs.harvard.edu/abs/2011ApJ...728...63R} {728, 63}

\bibitem[\protect\citeauthoryear{{Rasmussen} \& {Williams}}{{Rasmussen} \& {Williams}}{2006}]{Rasmussen2006}
{Rasmussen} C.~E.,  {Williams} C. K.~I.,  2006, {Gaussian Processes for Machine Learning}.
MIT press Cambridge, MA

\bibitem[\protect\citeauthoryear{Rasmussen, Bousquet, Luxburg  \& Rätsch}{Rasmussen et~al.}{2004}]{Rasmussen2004}
Rasmussen C.,  Bousquet O.,  Luxburg U.,   Rätsch G.,  2004, \mn@doi [Advanced Lectures on Machine Learning: ML Summer Schools 2003, Canberra, Australia, February 2 - 14, 2003, Tübingen, Germany, August 4 - 16, 2003, Revised Lectures, 63-71 (2004)] {10.1007/978-3-540-28650-9_4}, 3176

\bibitem[\protect\citeauthoryear{{Reguitti} et~al.,}{{Reguitti} et~al.}{2022}]{Reguitti2022}
{Reguitti} A.,  et~al., 2022, \mn@doi [\aap] {10.1051/0004-6361/202243340}, \href {https://ui.adsabs.harvard.edu/abs/2022A&A...662L..10R} {662, L10}

\bibitem[\protect\citeauthoryear{{Reynolds} et~al.,}{{Reynolds} et~al.}{2020}]{Reynolds2020}
{Reynolds} T.~M.,  et~al., 2020, \mn@doi [\mnras] {10.1093/mnras/staa365}, \href {https://ui.adsabs.harvard.edu/abs/2020MNRAS.493.1761R} {493, 1761}

\bibitem[\protect\citeauthoryear{{Richmond}, {Treffers}, {Filippenko}, {Paik}, {Leibundgut}, {Schulman}  \& {Cox}}{{Richmond} et~al.}{1994}]{Richmond1994}
{Richmond} M.~W.,  {Treffers} R.~R.,  {Filippenko} A.~V.,  {Paik} Y.,  {Leibundgut} B.,  {Schulman} E.,   {Cox} C.~V.,  1994, \mn@doi [\aj] {10.1086/116915}, \href {https://ui.adsabs.harvard.edu/abs/1994AJ....107.1022R} {107, 1022}

\bibitem[\protect\citeauthoryear{{Richmond} et~al.,}{{Richmond} et~al.}{1996}]{Richmond1996}
{Richmond} M.~W.,  et~al., 1996, \mn@doi [\aj] {10.1086/117785}, \href {https://ui.adsabs.harvard.edu/abs/1996AJ....111..327R} {111, 327}

\bibitem[\protect\citeauthoryear{{Riess}, {Press}  \& {Kirshner}}{{Riess} et~al.}{1996}]{Riess1996}
{Riess} A.~G.,  {Press} W.~H.,   {Kirshner} R.~P.,  1996, \mn@doi [\apj] {10.1086/178129}, \href {https://ui.adsabs.harvard.edu/abs/1996ApJ...473...88R} {473, 88}

\bibitem[\protect\citeauthoryear{{Robinson}}{{Robinson}}{2007}]{Robinson2007}
{Robinson} K.,  2007, {Spectroscopy: The Key to the Stars}.
Springer, \mn@doi{10.1007/978-0-387-68288-4}

\bibitem[\protect\citeauthoryear{{Roming} et~al.,}{{Roming} et~al.}{2005}]{Roming2005}
{Roming} P. W.~A.,  et~al., 2005, \mn@doi [\ssr] {10.1007/s11214-005-5095-4}, \href {https://ui.adsabs.harvard.edu/abs/2005SSRv..120...95R} {120, 95}

\bibitem[\protect\citeauthoryear{{Roming} et~al.,}{{Roming} et~al.}{2012}]{Roming2012}
{Roming} P.~W.~A.,  et~al., 2012, \mn@doi [\apj] {10.1088/0004-637X/751/2/92}, \href {https://ui.adsabs.harvard.edu/abs/2012ApJ...751...92R} {751, 92}

\bibitem[\protect\citeauthoryear{Ross et~al.,}{Ross et~al.}{2015}]{Ross2015}
Ross T.,  et~al., 2015, Central Bureau Electronic Telegrams, 4125, 1

\bibitem[\protect\citeauthoryear{{Roy} et~al.,}{{Roy} et~al.}{2011}]{Roy2011}
{Roy} R.,  et~al., 2011, \mn@doi [\apj] {10.1088/0004-637X/736/2/76}, \href {https://ui.adsabs.harvard.edu/abs/2011ApJ...736...76R} {736, 76}

\bibitem[\protect\citeauthoryear{{Rubin} et~al.,}{{Rubin} et~al.}{2016}]{Rubin2016}
{Rubin} A.,  et~al., 2016, \mn@doi [\apj] {10.3847/0004-637X/820/1/33}, \href {https://ui.adsabs.harvard.edu/abs/2016ApJ...820...33R} {820, 33}

\bibitem[\protect\citeauthoryear{{Sahu}, {Anupama}, {Srividya}  \& {Muneer}}{{Sahu} et~al.}{2006}]{Sahu2006}
{Sahu} D.~K.,  {Anupama} G.~C.,  {Srividya} S.,   {Muneer} S.,  2006, \mn@doi [\mnras] {10.1111/j.1365-2966.2006.10937.x}, \href {https://ui.adsabs.harvard.edu/abs/2006MNRAS.372.1315S} {372, 1315}

\bibitem[\protect\citeauthoryear{{Sahu}, {Tanaka}, {Anupama}, {Gurugubelli}  \& {Nomoto}}{{Sahu} et~al.}{2009}]{Sahu2009}
{Sahu} D.~K.,  {Tanaka} M.,  {Anupama} G.~C.,  {Gurugubelli} U.~K.,   {Nomoto} K.,  2009, \mn@doi [\apj] {10.1088/0004-637X/697/1/676}, \href {https://ui.adsabs.harvard.edu/abs/2009ApJ...697..676S} {697, 676}

\bibitem[\protect\citeauthoryear{{Sahu}, {Anupama}  \& {Chakradhari}}{{Sahu} et~al.}{2013}]{Sahu2013}
{Sahu} D.~K.,  {Anupama} G.~C.,   {Chakradhari} N.~K.,  2013, \mn@doi [\mnras] {10.1093/mnras/stt647}, \href {https://ui.adsabs.harvard.edu/abs/2013MNRAS.433....2S} {433, 2}

\bibitem[\protect\citeauthoryear{{Saunders} et~al.,}{{Saunders} et~al.}{2018}]{Saunders2018}
{Saunders} C.,  et~al., 2018, \mn@doi [\apj] {10.3847/1538-4357/aaec7e}, \href {https://ui.adsabs.harvard.edu/abs/2018ApJ...869..167S} {869, 167}

\bibitem[\protect\citeauthoryear{{Schlegel}, {Finkbeiner}  \& {Davis}}{{Schlegel} et~al.}{1998}]{Schlegel1998}
{Schlegel} D.~J.,  {Finkbeiner} D.~P.,   {Davis} M.,  1998, \mn@doi [\apj] {10.1086/305772}, \href {https://ui.adsabs.harvard.edu/abs/1998ApJ...500..525S} {500, 525}

\bibitem[\protect\citeauthoryear{{Shivvers} et~al.,}{{Shivvers} et~al.}{2017}]{Shivvers2017}
{Shivvers} I.,  et~al., 2017, \mn@doi [\pasp] {10.1088/1538-3873/aa54a6}, \href {https://ui.adsabs.harvard.edu/abs/2017PASP..129e4201S} {129, 054201}

\bibitem[\protect\citeauthoryear{{Shivvers} et~al.,}{{Shivvers} et~al.}{2019}]{Shivvers2019}
{Shivvers} I.,  et~al., 2019, \mn@doi [\mnras] {10.1093/mnras/sty2719}, \href {https://ui.adsabs.harvard.edu/abs/2019MNRAS.482.1545S} {482, 1545}

\bibitem[\protect\citeauthoryear{{Silverman} et~al.,}{{Silverman} et~al.}{2012}]{Silverman2012}
{Silverman} J.~M.,  et~al., 2012, \mn@doi [\mnras] {10.1111/j.1365-2966.2012.21270.x}, \href {https://ui.adsabs.harvard.edu/abs/2012MNRAS.425.1789S} {425, 1789}

\bibitem[\protect\citeauthoryear{{Silverman} et~al.,}{{Silverman} et~al.}{2017}]{Silverman2017}
{Silverman} J.~M.,  et~al., 2017, \mn@doi [\mnras] {10.1093/mnras/stx058}, \href {https://ui.adsabs.harvard.edu/abs/2017MNRAS.467..369S} {467, 369}

\bibitem[\protect\citeauthoryear{{Smartt} et~al.,}{{Smartt} et~al.}{2015}]{Smartt2015}
{Smartt} S.~J.,  et~al., 2015, \mn@doi [\aap] {10.1051/0004-6361/201425237}, \href {https://ui.adsabs.harvard.edu/abs/2015A&A...579A..40S} {579, A40}

\bibitem[\protect\citeauthoryear{{Sollerman} et~al.,}{{Sollerman} et~al.}{2021}]{Sollerman2021}
{Sollerman} J.,  et~al., 2021, \mn@doi [\aap] {10.1051/0004-6361/202141374}, \href {https://ui.adsabs.harvard.edu/abs/2021A&A...655A.105S} {655, A105}

\bibitem[\protect\citeauthoryear{{Sonbas} et~al.,}{{Sonbas} et~al.}{2008}]{Sonbas2008}
{Sonbas} E.,  et~al., 2008, \mn@doi [Astrophysical Bulletin] {10.1134/S1990341308030036}, \href {https://ui.adsabs.harvard.edu/abs/2008AstBu..63..228S} {63, 228}

\bibitem[\protect\citeauthoryear{{Srivastav}, {Anupama}  \& {Sahu}}{{Srivastav} et~al.}{2014}]{Srivastav2014}
{Srivastav} S.,  {Anupama} G.~C.,   {Sahu} D.~K.,  2014, \mn@doi [\mnras] {10.1093/mnras/stu1878}, \href {https://ui.adsabs.harvard.edu/abs/2014MNRAS.445.1932S} {445, 1932}

\bibitem[\protect\citeauthoryear{{Stritzinger} et~al.,}{{Stritzinger} et~al.}{2009}]{Stritzinger2009}
{Stritzinger} M.,  et~al., 2009, \mn@doi [\apj] {10.1088/0004-637X/696/1/713}, \href {https://ui.adsabs.harvard.edu/abs/2009ApJ...696..713S} {696, 713}

\bibitem[\protect\citeauthoryear{{Stritzinger} et~al.,}{{Stritzinger} et~al.}{2012}]{Stritzinger2012}
{Stritzinger} M.,  et~al., 2012, \mn@doi [\apj] {10.1088/0004-637X/756/2/173}, \href {https://ui.adsabs.harvard.edu/abs/2012ApJ...756..173S} {756, 173}

\bibitem[\protect\citeauthoryear{{Stritzinger} et~al.,}{{Stritzinger} et~al.}{2018}]{Stritzinger2018}
{Stritzinger} M.~D.,  et~al., 2018, \mn@doi [\aap] {10.1051/0004-6361/201730842}, \href {https://ui.adsabs.harvard.edu/abs/2018A&A...609A.134S} {609, A134}

\bibitem[\protect\citeauthoryear{{Sutherland} \& {Wheeler}}{{Sutherland} \& {Wheeler}}{1984}]{Sutherland1984}
{Sutherland} P.~G.,  {Wheeler} J.~C.,  1984, \mn@doi [\apj] {10.1086/161995}, \href {https://ui.adsabs.harvard.edu/abs/1984ApJ...280..282S} {280, 282}

\bibitem[\protect\citeauthoryear{{Taddia} et~al.,}{{Taddia} et~al.}{2013}]{Taddia2013}
{Taddia} F.,  et~al., 2013, \mn@doi [\aap] {10.1051/0004-6361/201321180}, \href {https://ui.adsabs.harvard.edu/abs/2013A&A...555A..10T} {555, A10}

\bibitem[\protect\citeauthoryear{{Taddia} et~al.,}{{Taddia} et~al.}{2016}]{Taddia2016}
{Taddia} F.,  et~al., 2016, \mn@doi [\aap] {10.1051/0004-6361/201628703}, \href {https://ui.adsabs.harvard.edu/abs/2016A&A...592A..89T} {592, A89}

\bibitem[\protect\citeauthoryear{{Taddia}, {Sollerman}, {Fremling}, {Karamehmetoglu}, {Barbarino}, {Lunnan}, {West}  \& {Gal-Yam}}{{Taddia} et~al.}{2019a}]{Taddia2019b}
{Taddia} F.,  {Sollerman} J.,  {Fremling} C.,  {Karamehmetoglu} E.,  {Barbarino} C.,  {Lunnan} R.,  {West} S.,   {Gal-Yam} A.,  2019a, \mn@doi [\aap] {10.1051/0004-6361/201833688}, \href {https://ui.adsabs.harvard.edu/abs/2019A&A...621A..64T} {621, A64}

\bibitem[\protect\citeauthoryear{{Taddia} et~al.,}{{Taddia} et~al.}{2019b}]{Taddia2019}
{Taddia} F.,  et~al., 2019b, \mn@doi [\aap] {10.1051/0004-6361/201834429}, \href {https://ui.adsabs.harvard.edu/abs/2019A&A...621A..71T} {621, A71}

\bibitem[\protect\citeauthoryear{{Takaki} et~al.,}{{Takaki} et~al.}{2013}]{Takaki2013}
{Takaki} K.,  et~al., 2013, \mn@doi [\apjl] {10.1088/2041-8205/772/2/L17}, \href {https://ui.adsabs.harvard.edu/abs/2013ApJ...772L..17T} {772, L17}

\bibitem[\protect\citeauthoryear{{Tak{\'a}ts}}{{Tak{\'a}ts}}{2014}]{Takats2014}
{Tak{\'a}ts} K.,  2014, in Revista Mexicana de Astronomia y Astrofisica Conference Series. pp 169--169

\bibitem[\protect\citeauthoryear{{Tak{\'a}ts} et~al.,}{{Tak{\'a}ts} et~al.}{2015}]{Takats2015}
{Tak{\'a}ts} K.,  et~al., 2015, \mn@doi [\mnras] {10.1093/mnras/stv857}, \href {https://ui.adsabs.harvard.edu/abs/2015MNRAS.450.3137T} {450, 3137}

\bibitem[\protect\citeauthoryear{{Tartaglia} et~al.,}{{Tartaglia} et~al.}{2015}]{Tartaglia2015}
{Tartaglia} L.,  et~al., 2015, The Astronomer's Telegram, \href {https://ui.adsabs.harvard.edu/abs/2015ATel.8039....1T} {8039, 1}

\bibitem[\protect\citeauthoryear{{Tartaglia} et~al.,}{{Tartaglia} et~al.}{2017}]{Tartaglia2017}
{Tartaglia} L.,  et~al., 2017, \mn@doi [\apjl] {10.3847/2041-8213/aa5c7f}, \href {https://ui.adsabs.harvard.edu/abs/2017ApJ...836L..12T} {836, L12}

\bibitem[\protect\citeauthoryear{{Taubenberger} et~al.,}{{Taubenberger} et~al.}{2009}]{Taubenberger2009}
{Taubenberger} S.,  et~al., 2009, \mn@doi [\mnras] {10.1111/j.1365-2966.2009.15003.x}, \href {https://ui.adsabs.harvard.edu/abs/2009MNRAS.397..677T} {397, 677}

\bibitem[\protect\citeauthoryear{{Taubenberger} et~al.,}{{Taubenberger} et~al.}{2011}]{Taubenberger2011}
{Taubenberger} S.,  et~al., 2011, \mn@doi [\mnras] {10.1111/j.1365-2966.2011.18287.x}, \href {https://ui.adsabs.harvard.edu/abs/2011MNRAS.413.2140T} {413, 2140}

\bibitem[\protect\citeauthoryear{{Taubenberger} et~al.,}{{Taubenberger} et~al.}{2015}]{Taubenberger2015}
{Taubenberger} S.,  et~al., 2015, \mn@doi [\mnras] {10.1093/mnrasl/slu201}, \href {https://ui.adsabs.harvard.edu/abs/2015MNRAS.448L..48T} {448, L48}

\bibitem[\protect\citeauthoryear{{Teffs}, {Prentice}, {Mazzali}  \& {Ashall}}{{Teffs} et~al.}{2021}]{Teffs2021}
{Teffs} J.~J.,  {Prentice} S.~J.,  {Mazzali} P.~A.,   {Ashall} C.,  2021, \mn@doi [\mnras] {10.1093/mnras/stab258}, \href {https://ui.adsabs.harvard.edu/abs/2021MNRAS.502.3829T} {502, 3829}

\bibitem[\protect\citeauthoryear{{Terreran} et~al.,}{{Terreran} et~al.}{2016}]{Terreran2016}
{Terreran} G.,  et~al., 2016, \mn@doi [\mnras] {10.1093/mnras/stw1591}, \href {https://ui.adsabs.harvard.edu/abs/2016MNRAS.462..137T} {462, 137}

\bibitem[\protect\citeauthoryear{{Terreran} et~al.,}{{Terreran} et~al.}{2019}]{Terreran2019}
{Terreran} G.,  et~al., 2019, \mn@doi [\apj] {10.3847/1538-4357/ab3e37}, \href {https://ui.adsabs.harvard.edu/abs/2019ApJ...883..147T} {883, 147}

\bibitem[\protect\citeauthoryear{{Terreran} et~al.,}{{Terreran} et~al.}{2022}]{Terreran2022}
{Terreran} G.,  et~al., 2022, \mn@doi [\apj] {10.3847/1538-4357/ac3820}, \href {https://ui.adsabs.harvard.edu/abs/2022ApJ...926...20T} {926, 20}

\bibitem[\protect\citeauthoryear{{Tinyanont} et~al.,}{{Tinyanont} et~al.}{2022}]{Tinyanont2022}
{Tinyanont} S.,  et~al., 2022, \mn@doi [\mnras] {10.1093/mnras/stab2887}, \href {https://ui.adsabs.harvard.edu/abs/2022MNRAS.512.2777T} {512, 2777}

\bibitem[\protect\citeauthoryear{{Tomasella} et~al.,}{{Tomasella} et~al.}{2013}]{Tomasella2013}
{Tomasella} L.,  et~al., 2013, \mn@doi [\mnras] {10.1093/mnras/stt1130}, \href {https://ui.adsabs.harvard.edu/abs/2013MNRAS.434.1636T} {434, 1636}

\bibitem[\protect\citeauthoryear{{Tomasella} et~al.,}{{Tomasella} et~al.}{2018}]{Tomasella2018}
{Tomasella} L.,  et~al., 2018, \mn@doi [\mnras] {10.1093/mnras/stx3220}, \href {https://ui.adsabs.harvard.edu/abs/2018MNRAS.475.1937T} {475, 1937}

\bibitem[\protect\citeauthoryear{{Trumpler}}{{Trumpler}}{1930}]{Trumpler1930}
{Trumpler} R.~J.,  1930, \mn@doi [\pasp] {10.1086/124039}, \href {https://ui.adsabs.harvard.edu/abs/1930PASP...42..214T} {42, 214}

\bibitem[\protect\citeauthoryear{{Tsvetkov}, {Volkov}, {Sorokina}, {Blinnikov}, {Pavlyuk}  \& {Borisov}}{{Tsvetkov} et~al.}{2012}]{Tsvetkov2012}
{Tsvetkov} D.~Y.,  {Volkov} I.~M.,  {Sorokina} E.~I.,  {Blinnikov} S.~I.,  {Pavlyuk} N.~N.,   {Borisov} G.~V.,  2012, VizieR Online Data Catalog (other), \href {https://ui.adsabs.harvard.edu/abs/2012yCatp012003201T} {0120, J/other/PZ/32}

\bibitem[\protect\citeauthoryear{\VAN{Dyk}{Van}{Van}~Dyk et~al.,}{\VAN{Dyk}{Van}{Van}~Dyk et~al.}{2014}]{VanDyk2014}
\VAN{Dyk}{Van}{Van}~Dyk S.~D.,  et~al., 2014, \mn@doi [\aj] {10.1088/0004-6256/147/2/37}, \href {https://ui.adsabs.harvard.edu/abs/2014AJ....147...37V} {147, 37}

\bibitem[\protect\citeauthoryear{{Valenti} et~al.,}{{Valenti} et~al.}{2008}]{Valenti2008}
{Valenti} S.,  et~al., 2008, \mn@doi [\apjl] {10.1086/527672}, \href {https://ui.adsabs.harvard.edu/abs/2008ApJ...673L.155V} {673, L155}

\bibitem[\protect\citeauthoryear{{Valenti} et~al.,}{{Valenti} et~al.}{2011}]{Valenti2011}
{Valenti} S.,  et~al., 2011, \mn@doi [\mnras] {10.1111/j.1365-2966.2011.19262.x}, \href {https://ui.adsabs.harvard.edu/abs/2011MNRAS.416.3138V} {416, 3138}

\bibitem[\protect\citeauthoryear{{Valenti} et~al.,}{{Valenti} et~al.}{2012}]{Valenti2012}
{Valenti} S.,  et~al., 2012, \mn@doi [\apjl] {10.1088/2041-8205/749/2/L28}, \href {https://ui.adsabs.harvard.edu/abs/2012ApJ...749L..28V} {749, L28}

\bibitem[\protect\citeauthoryear{{Valenti} et~al.,}{{Valenti} et~al.}{2014a}]{Valenti2014b}
{Valenti} S.,  et~al., 2014a, \mn@doi [\mnras] {10.1093/mnras/stt1983}, \href {https://ui.adsabs.harvard.edu/abs/2014MNRAS.437.1519V} {437, 1519}

\bibitem[\protect\citeauthoryear{{Valenti} et~al.,}{{Valenti} et~al.}{2014b}]{Valenti2014}
{Valenti} S.,  et~al., 2014b, \mn@doi [\mnras] {10.1093/mnrasl/slt171}, \href {https://ui.adsabs.harvard.edu/abs/2014MNRAS.438L.101V} {438, L101}

\bibitem[\protect\citeauthoryear{{Valenti} et~al.,}{{Valenti} et~al.}{2015}]{Valenti2015}
{Valenti} S.,  et~al., 2015, \mn@doi [\mnras] {10.1093/mnras/stv208}, \href {https://ui.adsabs.harvard.edu/abs/2015MNRAS.448.2608V} {448, 2608}

\bibitem[\protect\citeauthoryear{{Valenti} et~al.,}{{Valenti} et~al.}{2016}]{Valenti2016}
{Valenti} S.,  et~al., 2016, \mn@doi [\mnras] {10.1093/mnras/stw870}, \href {https://ui.adsabs.harvard.edu/abs/2016MNRAS.459.3939V} {459, 3939}

\bibitem[\protect\citeauthoryear{{Valerin} et~al.,}{{Valerin} et~al.}{2022}]{Valerin2022}
{Valerin} G.,  et~al., 2022, \mn@doi [\mnras] {10.1093/mnras/stac1182}, \href {https://ui.adsabs.harvard.edu/abs/2022MNRAS.513.4983V} {513, 4983}

\bibitem[\protect\citeauthoryear{{Vincenzi}, {Sullivan}, {Firth}, {Guti{\'e}rrez}, {Frohmaier}, {Smith}, {Angus}  \& {Nichol}}{{Vincenzi} et~al.}{2019}]{Vincenzi2019}
{Vincenzi} M.,  {Sullivan} M.,  {Firth} R.~E.,  {Guti{\'e}rrez} C.~P.,  {Frohmaier} C.,  {Smith} M.,  {Angus} C.,   {Nichol} R.~C.,  2019, \mn@doi [\mnras] {10.1093/mnras/stz2448}, \href {https://ui.adsabs.harvard.edu/abs/2019MNRAS.489.5802V} {489, 5802}

\bibitem[\protect\citeauthoryear{Virtanen et~al.,}{Virtanen et~al.}{2020}]{Virtanen2020}
Virtanen P.,  et~al., 2020, \mn@doi [Nature Methods] {10.1038/s41592-019-0686-2}, \href {https://rdcu.be/b08Wh} {17, 261}

\bibitem[\protect\citeauthoryear{{Walker} et~al.,}{{Walker} et~al.}{2014}]{Walker2014}
{Walker} E.~S.,  et~al., 2014, \mn@doi [\mnras] {10.1093/mnras/stu1017}, \href {https://ui.adsabs.harvard.edu/abs/2014MNRAS.442.2768W} {442, 2768}

\bibitem[\protect\citeauthoryear{{Wheeler}, {Johnson}  \& {Clocchiatti}}{{Wheeler} et~al.}{2015}]{Wheeler2015}
{Wheeler} J.~C.,  {Johnson} V.,   {Clocchiatti} A.,  2015, \mn@doi [\mnras] {10.1093/mnras/stv650}, \href {https://ui.adsabs.harvard.edu/abs/2015MNRAS.450.1295W} {450, 1295}

\bibitem[\protect\citeauthoryear{{Whitelock} et~al.,}{{Whitelock} et~al.}{1988}]{Whitelock1988}
{Whitelock} P.~A.,  et~al., 1988, \mn@doi [\mnras] {10.1093/mnras/234.1.5P}, \href {https://ui.adsabs.harvard.edu/abs/1988MNRAS.234P...5W} {234, 5P}

\bibitem[\protect\citeauthoryear{{Whitford}}{{Whitford}}{1958}]{Whitford1958}
{Whitford} A.~E.,  1958, \mn@doi [\aj] {10.1086/107725}, \href {https://ui.adsabs.harvard.edu/abs/1958AJ.....63..201W} {63, 201}

\bibitem[\protect\citeauthoryear{Woosley}{Woosley}{1988}]{Woosley1988}
Woosley S.,  1988, Astrophysical Journal, Part 1 (ISSN 0004-637X), vol. 330, July 1, 1988, p. 218-253., 330, 218

\bibitem[\protect\citeauthoryear{{Woosley}, {Langer}  \& {Weaver}}{{Woosley} et~al.}{1993}]{Woosley1993}
{Woosley} S.~E.,  {Langer} N.,   {Weaver} T.~A.,  1993, \mn@doi [\apj] {10.1086/172886}, \href {https://ui.adsabs.harvard.edu/abs/1993ApJ...411..823W} {411, 823}

\bibitem[\protect\citeauthoryear{{Yang} et~al.,}{{Yang} et~al.}{2021}]{Yang2021}
{Yang} S.,  et~al., 2021, \mn@doi [\aap] {10.1051/0004-6361/202141244}, \href {https://ui.adsabs.harvard.edu/abs/2021A&A...655A..90Y} {655, A90}

\bibitem[\protect\citeauthoryear{{Yao} et~al.,}{{Yao} et~al.}{2020}]{Yao2020}
{Yao} Y.,  et~al., 2020, \mn@doi [\apj] {10.3847/1538-4357/abaa3d}, \href {https://ui.adsabs.harvard.edu/abs/2020ApJ...900...46Y} {900, 46}

\bibitem[\protect\citeauthoryear{{Yaron} \& {Gal-Yam}}{{Yaron} \& {Gal-Yam}}{2012}]{Yaron2012}
{Yaron} O.,  {Gal-Yam} A.,  2012, \mn@doi [\pasp] {10.1086/666656}, \href {https://ui.adsabs.harvard.edu/abs/2012PASP..124..668Y} {124, 668}

\bibitem[\protect\citeauthoryear{{Yaron} et~al.,}{{Yaron} et~al.}{2017}]{Yaron2017}
{Yaron} O.,  et~al., 2017, \mn@doi [Nature Physics] {10.1038/nphys4025}, \href {https://ui.adsabs.harvard.edu/abs/2017NatPh..13..510Y} {13, 510}

\bibitem[\protect\citeauthoryear{{Yuan} et~al.,}{{Yuan} et~al.}{2016}]{Yuan2016}
{Yuan} F.,  et~al., 2016, \mn@doi [\mnras] {10.1093/mnras/stw1419}, \href {https://ui.adsabs.harvard.edu/abs/2016MNRAS.461.2003Y} {461, 2003}

\bibitem[\protect\citeauthoryear{{Zhang} et~al.,}{{Zhang} et~al.}{2014}]{Zhang2014}
{Zhang} J.,  et~al., 2014, \mn@doi [\apj] {10.1088/0004-637X/797/1/5}, \href {https://ui.adsabs.harvard.edu/abs/2014ApJ...797....5Z} {797, 5}

\makeatother
\end{thebibliography}


\begin{table*}
\centering
\caption{Catalogue of CCSNe employed in \texttt{CASTOR}. For each supernova we specified: the spectral class and the redshift as catalogued on WISeREP \protect\citep{Yaron2012}, the available filters, the number of selected light curves and spectra and the references}
\label{tab:trainingset}
\begin{tabular}{lcccccr}

\hline
Name & Class & Redshift & Filters & Light Curves & Spectra & References* \\ 
\hline

iPTF13bvn	&	Ib	    &	0.00449	&	U, B, V, R, I, g, r, i, z, w2, w1, m2	&	12	&	30  & (1) \\ 	
PTF10qts	&	Ic-BL	&	0.0907	&	B, R, g, r, i, z	    &	6	&	6	                            & (2) \\ 
ASASSN14jb	&	II-P	&	0.006	&	B, V, g, r, i 	        &	5	&	10	                            & (3) \\ 
ASASSN15ed	&	Ib	    &	0.04866	&	B, V, u, g, r, i, z	&	7	&	15	                            & (4) \\ 
ASASSN15oz	&	II	    &	0.007	&	U, B, V, R, I, u, g, r, i, w2, w1, m2	&	12	&	17	& (5) \\ 
SN1987A	    &	II	    &	0.001067&	U, B, V, R, I 	&	5	&	18	                                    & (6) \\	
SN1993J	    &	II-b	&	-0.000113&	U, B, V, R, I 	&	5	&	52	                                    & (7) \\ 	
SN1994I	    &	Ic	    &	0.001544 &	U, B, V, R, I 	&	5	&	36	                                    & (8), (9) \\ 	
SN1998bw	&	Ic	    &	0.008499 &	U, B, V, R, I 	&	5	&	25	                                    & (10) \\	
SN1999dn	&	Ib	    &	0.009366 &	U, B, V, R, I 	&	5	&	12	                                    & (11) \\	
SN1999em	&	II-P	&	0.002392 &	U, B, V, R, I 	&	5	&	32	                                    & (12) \\ 	
SN2002ap	&	Ic-BL	&	0.002108 &	U, B, V, R, I 	&	5	&	30	                                    & (9), (13) \\ 	
SN2004aw	&	Ic	    &	0.0163 &	U, B, V, R, I 	&	5	&	26	                                    & (14) \\ 
SN2004et	&	II-P	&	0.00016	&	U, B, V, R, I 	&	5	&	24	                                    & (15) \\ 
SN2004fe	&	Ic	    & 	0.018	&	B, V, R, u, g, r, i, r’, i’,	&	9	&	13	                & (9), (16), (17) \\ 
SN2004gq	&	Ib	    &	0.006468	&	U, B, V, R, u, g, r, i, r’, i’, Y, J, H	&	13	&	23	& (9), (17), (18) \\ 
SN2004gt	&	Ic	    &	0.005477	&	B, V, R, u, g, r, i, r', i', Y, J, H, Ks	&	13	&	17	& (9), (17), (19), (20) \\ 
SN2004gv	&	Ib	    &	0.02	&	B, V, R, u, g, r, i, Y, J, H	&	10	&	6	                    & (9), (17), (20) \\ 
SN2005bf	&	Ib/c	&	0.018913	&	U, B, V, u, g, r, i, r', i', J, H, Ks, K	&	13	&	29	& (9), (17), (20), (21) \\ 
SN2005cs	&	II-P	&	0.002	&	U, B, V, R, I, z, J, H, K, w2, w1, m2	&	12	&	29	        & (22) \\ 
SN2005hg	&	Ib	    &	0.021308	&	U, B, V, R, r', i', J, H, Ks	&	9	&	20	                    & (9), (20), (23) \\ 
SN2005ip	&	II-n	&	0.007	&	B, V, u, g, r, i, Y, J, H	&	9	&	13	                            & (24) \\ 
SN2005kj	&	II-n	&	0.017	&	B, V, u, g, r, i, Y, J, H, Ks	&	10	&	15	                    & (25) \\ 
SN2006aa	&	II-n	&	0.0207	&	B, V, u, g, r, i, Y, J, H	&	9	&	7	                            & (25) \\ 
SN2006aj	&	Ic	&	0.033023	&	U, B, V, r', i', J, H, Ks, w2, w1, m2	&	11	&	19	            & (9), (23), (26) \\ 
SN2006ep	&	Ib	&	0.015134	&	B, V, u, g, r, i, r', i', Y, J, H	&	11	&	10	                & (9), (17), (20) \\ 
SN2006jd	&	II-n	&	0.0185 &	B, V, u, g, r, i, Y, J, H	&	9	&	10	                            & (24) \\ 
SN2006qq	&	II-n	&	0.029	&	B, V, u, g, r, i, Y, J, H	&	9	&	9	                            & (25) \\ 
SN2006T	    &	II-b	&	0.008091	&	U, B, V, u, g, r, i, r', i', Y, J, H	&	12	&	17	    & (9), (17), (18) \\ 
SN2007gr	&	Ic	&	0.001728	&	U, B, V, R, I, r', i', J, H, Ks	&	10	&	35	                    & (9), (20), (23), (27) \\ 
SN2007od	&	II	&	0.00586	&	U, B, V, R, I, r', i', w2, w1, m2	&	10	&	15	                    & (28) \\ 
SN2007pk	&	II-n	&	0.016655	&	U, B, V, R, I, r', i', w2, w1, m2	&	10	&	9	            & (29), (30) \\ 
SN2007ru	&	Ic-BL	&	0.015464	&	U, B, V, R, I, r', i'	&	7	&	5	                                & (9), (20), (31) \\ 
SN2007uy	&	Ib	&	0.007	&	U, B, V, r', i', J, H, Ks, w2, w1, m2	&	11	&	12	                & (9), (20), (23) \\ 
SN2007Y	    &	Ib/c	&	0.004657	&	U, B, V, u, g, r, i, Y, J, H, w2, w1, m2	&	13	&	9	& (17), (20), (32) \\ 
SN2008aq	&	II-b	&	0.007972	&	U, B, V, u, g, r, i, r', i', Y, J, H, w2, w1, m2	&	15	&	19	        & (9), (17) \\ 
SN2008ax	&	II-b	&	0.001931	&	U, B, V, R, I, u', g', r', i', z', J, H, Ks, w2, w1, m2	&	16	&	48	& (9), (23), (33) \\ 
SN2008bj	&	II	    &	0.019	&	U, B, V, r', i'	&	5	&	16	                                               &  (30) \\   
SN2008bo	&	II-b	&	0.00495	&	U, B, V, r', i', w2, w1, m2	&	8	&	17	                               & (9), (20), (23) \\ 
SN2008D	    &	Ib	    &	0.006521	&	U, B, V, R, I, r', i', J, H, Ks, w2, w1, m2 	&	13	&	30 & (9), (20), (23), (34)  \\ 	
SN2008fq	&	II-n	&	0.0106 &	B, V, u, g, r, i, Y, J, H, Ks	&	10	&	10	                       & (25), (35) \\ 
SN2008in	&	II-P	&	0.005224	&	U, B, V, R, I, r', i', J, H, K, w2, w1, m2 	&	13	&	12 & (30), (36) \\ 	
SN2009bb	&	Ic-BL	&	0.0104 	&	B, V, R, I, u, g, r, i, Y, J, H, w1 	&	12	&	15	           & (17), (37) \\ 
SN2009bw	&	II-P	&	0.0039	&	U, B, V, R, I	&	5	&	16	                                                   & (38) \\ 
SN2009dd	&	II	&	0.0034	&	U, B, V, R, I, r', i' 	&	7	&	13	                                       & (29), (30) \\ 
SN2009ib	&	II-P	&	0.00435	&	B, V, R, I, u', g', r', i', z', J, H 	&	11	&	10	               & (39) \\ 
SN2009ip	&	II-n	&	0.005944	&	U, B, V, R, I, J, H, w2, w1, m2 	&	10	&	26	               & (40), (41), (42) \\ 
SN2009iz	&	Ib	&	0.014196	&	U, B, V, u', r', i', J, H, Ks, w2, w1 	&	11	&	11	           & (9), (20), (23) \\ 
SN2009jf	&	Ib	&	0.008	&	U, B, V, R, I, u, g, r, i, z, J, H, Ks 	&	13	&	34	           & (9), (20), (23), (43) \\ 
SN2009kr	&	II	&	0.0065	&	U, B, V, R, I, J, H, K, w2, w1, m2 	&	11	&	9	                   & (44) \\

\hline
\end{tabular}
\end{table*}

\begin{table*}
\centering
\contcaption{}
\label{tab:continued}

\begin{tabular}{lcccccr} 
\hline
Name & Class & Redshift & Filters & Light Curves & Spectra & References* \\ 
\hline

SN2009N	    &	II-P	&	0.003449	&	U, B, V, R, I, g', r', i', z', w2, w1, m2 	&	12	&	25	   & (45) \\ 	
SN2010al	&	Ib/n	&	0.017155	&	U, B, V, u', r', i', J, H, K, w2, w1, m2 	&	12	&	17	   & (46) \\ 	
SN2011bm	&	Ic	&	0.022	&	U, B, V, R, I, u, g, r, i, z 	&	10	&	16	                           & (47) \\	
SN2011dh	&	II-b	&	0.002	&	U, B, V, R, I, u, g, r, i, z, w2, w1, m2 	&	13	&	44	       & (48) \\ 	
SN2011ei	&	II-b	&	0.009313	&	U, B, V, R, I, u, w2, w1, m2 	&	9	&	14	                       & (49) \\ 	
SN2011fu	&	II-b	&	0.018489	&	U, B, V, R, I, u, z, J, H, Ks 	&	10	&	32	                   & (20), (50) \\ 	
SN2011hs	&	II-b	&	0.0057	&	U, B, V, R, I, u', g', r', i', z', w2, w1, m2 	&	13	&	14 & (51) \\		
SN2011ht	&	II-n	&	0.0036	&	U, B, V, R, I, J, H, K, w2, w1, m2 	&	11	&	10	               & (52) \\ 	
SN2011kn	&	Ic	&	0.0603	&	U, R, g, r, i 	&	5	&	5	                                                   & (53), (54) \\	
SN2011ko	&	Ic	&	0.0744	&	U, R, g, r, i 	&	5	&	5	                                                   & (53), (55) \\ 	
SN2012A	&	II-P	&	0.002512	&	U, B, V, R, I, g, r, i, z, w2, w1, m2 	&	12	&	6	           & (56) \\ 	
SN2012ap	&	Ib/c	&	0.01224	&	U, B, V, R, I, w2, w1, m2 	&	8	&	11	                               & (20), (57) \\ 	
SN2012au	&	Ib	&	0.0045	&	U, B, V, w2, w1, m2 	&	6	&	7	                                           & (20), (58) \\ 	
SN2012aw	&	II-P	&	0.002595	&	U, B, V, R, I, u, g, r, i, z, J, H, K, w2, w1, m2 	&	16	&	40	& (59), (60) \\ 	
SN2012ec	&	II-P	&	0.0047	&	B, V, R, I, u', g', r', i', z', J, H, K 	&	12	&	30	        & (42), (61) \\ 	
SN2012hn	&	Ic-Ca-rich	&	0.0076	&	B, V, R, I, u, g 	&	6	&	9	                                        & (62) \\ 	
SN2013ab	&	II-P	&	0.0046	&	U, B, V, R, I, g, r, i, w2, w1, m2 	&	11	&	41	                & (63) \\ 	
SN2013ai	&	II	&	0.009	&	B, V, R, I, J, H 	&	6	&	13	                                                & (41), (42), (64) \\ 	
SN2013am	&	II-P	&	0.003	&	U, B, V, R, I, g, r, i,  J, H, K,  w2, w1, m2 	&	14	&	19	& (41), (65), (66) \\ 	
SN2013bb	&	II-b	&	0.019	&	B, V, g, r, i 	&	5	&	5	                                                & (67) \\	
SN2013by	&	II	&	0.003816	&	U, B, V, u, g, r, i, w2, w1, m2 	&	10	&	8	                    & (42), (68) \\ 	
SN2013df	&	II-b	&	0.002395	&	U, B, V, R, I, u, r, i, w2, w1, m2 	&	11	&	20	            & (20), (69), (70) \\ 	
SN2013ej	&	II	&	0.002192	&	U, B, V, R, I, u, g, r, i, z, J, H, Ks, K, w2, w1, m2 	&	17	&	29	& (71) \\ 	
SN2013fc	&	II-n	&	0.018	&	U, B, V, R, I, g, r, i, J, H, K 	&	11	&	17	                    & (72) \\ 	
SN2013fs	&	II-P	&	0.011855	&	U, B, V, R, I, g, r, i, z, w2, w1, m2 	&	12	&	24	        & (42), (60), (73) \\ 	
SN2013ge	&	Ib/c	&	0.004356	&	U, B, V, R, I, r, i, w2, w1, m2 	&	10	&	33	                & (74) \\ 	
SN2013K	&	II-P	&	0.008066	&	U, B, V, R, I, g, r, i, J, H, K 	&	11	&	12	                    & (41), (66) \\ 	
SN2014G	&	II-L	&	0.0039	&	U, B, V, R, I, u, g, r, i, z, w2, w1, m2 	&	13	&	28	            &	(70), (75) \\ 
SN2015ah	&	Ib	&	0.016	&	B, V, u, g, r, i, z, J, H, K 	&	10	&	30	                            & (20), (41), (67) \\  	
SN2015fn	&	Ic	&	0.054	&	U, B, g, r, i 	&	5	&	10	                                                    &	(53), (76) \\ 
SN2016aqf	&	II	&	0.004	&	B, V, g, r, i 	&	5	&	16	&	(53)  \\
SN2016bdu	&	II-n	&	0.017	&	B, V, u, g, r, i, z, J, H, K 	&	10	&	13	& (77) \\ 	
SN2016bkv	&	II	&	0.002	&	U, B, V, R, I, g, r, i, J, H, Ks 	&	11	&	16	&	(78) \\ 
SN2016coi	&	Ic-BL	&	0.0036	&	U, B, V, R, I, u, g, r, i, z, J, H, K 	&	13	&	13	& (79) \\ 	
SN2016flq	&	Ic	&	0.06	&	R, u, g, r, i 	&	5	&	8	&	(53) \\ 
SN2016gkg	&	II-b	&	0.0049	&	U, B, V, R, I, g, r, i, w2, w1, m2 	&	11	&	17	&	(80) \\  
SN2016gsd	&	II	&	0.06	&	U, B, V, g, r, i, z 	&	7	&	9	&	(41), (81) \\ 
SN2016iae	&	Ic	&	0.004	&	B, V, g, r, i, J, H, K 	&	8	&	14	&	(41), (67) \\ 
SN2016jdw	&	Ib	&	0.0189	&	u, g, r, i, z 	&	5	&	13	&	(41), (67) \\ 
SN2016P	&	Ic-BL	&	0.0146 &	B, V, u, g, r, i, z 	&	7	&	10	&	(41), (67) \\ 
SN2016X	&	II-P	&	0.0044	&	U, B, V, R, I, g, r, i, w2, w1, m2 	&	11	&	21	& (82) \\ 	
SN2017bgu	&	Ib	&	0.008	&	u, g, r, i, z 	&	5	&	13	&	(41), (67) \\ 
SN2017dcc	&	Ic-BL	&	0.0245 &	g, r, i, z, J, H, K 	&	7	&	5	&	(41), (67) \\ 
SN2017dio	&	Ic	&	0.037	&	u, g, r, i, z 	&	5	&	10	&	(83) \\ 
SN2017ein	&	Ic	&	0.0027	&	u, g, r, i, z 	&	5	&	18	&	(84) \\ 
SN2017hyh	&	II-b	&	0.012	&	B, V, g, r, i 	&	5	&	6	&	(41), (67) \\ 
SN2018bsz	&	II	&	0.026665	&	g, r, i, U, B, V, J, H, K&	9	&	16	&	(41), (85) \\ 	
SN2018cow	&	Ic-BL	&	0.014	&	u, g, r, i, z, g', r', i', z', J, H, K 	&	12	&	15	&	(86) \\ 
SN2018ec	&	Ic 	&	0.009	&	g, r, i, z, J, H 	&	6	&	7	&	(41), (87) \\ 
SN2018kzr	&	Ic 	&	0.056	&	B, g, r, i, z, w2, w1, m2 	&	8	&	5	&	(88) \\

\hline
\end{tabular}
\end{table*}

\begin{table*}
\centering
\contcaption{}
\label{tab:continued2}
\begin{tabular}{lcccccr} 
\hline
Name & Class & Redshift & Filters & Light Curves & Spectra & References* \\ 
\hline
																		
SN2019bkc	&	Ic-Ca-rich	&	0.04	&	B, V, R, I, g, r, i, z 	&	8	&	11	&	(41), (90) \\ 
SN2019dge	&	Ib	&	0.0213	&	u, g, r, i, z 	&	5	&	7	&	(91) \\ 
SN2020cxd	&	II	&	0.0039	&	U, B, V, g, r, i, z 	&	7	&	10	&	(92), (93), (94) \\ 
SN2020fkb	&   Ib	&   0.01041	&   B, V, u, g, r, i, z, J, H, K 	& 10	& 7 & (41), (87) \\  							
SN2020fqv	&	II	&	0.0075	&	U, B, V, g, r, i, z 	&	7	&	19	&	(94), (95) \\ 
SN2020jfo	&	II	&	0.0052	&	U, B, V, g, r, i, z, w2, w1, m2 	&	10	&	26	&	(94), (96) \\ 
SN2020pni	&	II	&	0.0169	&	U, B, V, u, g, r, i, z, w2, w1, m2 	&	11	&	18	&	(94), (97) \\ 
SN2021aai	&	II	&	0.007412	&	B, V, u, g, r, i, z, J, H 	&	9	&	8	&	(93) \\ 
SN2021foa	&	II-n	&	0.0084	&	U, B, u, g, r, i, z, w2, w1, m2 	&	10	&	16	&	(98) \\ 
SN2021yja	&	II	&	0.005307	&	U, g, r, i, w2, w1, m2	&	7	&	47	&	(94), (99) \\

\hline
\\
\multicolumn{6}{c}{
\begin{minipage} {12.7cm}
\textit{References}: (1) \citet{Cao2013, Fremling2014, Fremling2016, Srivastav2014}; (2) \citet{Walker2014, Taddia2019}; (3) \citet{Meza2019}; (4) \citet{Pastorello2015b}; (5) \citet{Bostroem2019}; (6) \citet{Catchpole1987,  Catchpole1988, Catchpole1989, Whitelock1988, Pun1995}; (7) \citet{Richmond1994, Barbon1995, Matheson2000a, Matheson2000b}; (8) \citet{Filippenko1995, Clocchiatti1996, Richmond1996}; (9) \citet{Modjaz2014}; (10) \citet{Galama1998, Patat2001}; (11) \citet{Benetti2011}; (12) \citet{Hamuy2001, Leonard2002, Galbany2016}; (13) \citet{Gal-Yam2002, Foley2003, Pandey2003, Chornock2013}; (14) \citet{Taubenberger2015}; (15) \citet{Sahu2006, Maguire2010, Faran2014a}; (16) \citet{Harutyunyan2008}; (17) \citet{Stritzinger2018}; (18) \citet{Shivvers2017}; (19) \citet{Modjaz2008, Taubenberger2009}; (20) \citet{Shivvers2019}; (21) \citet{Folatelli2006}; (22) \citet{Pastorello2006, Pastorello2009, Bufano2009}; (23) \citet{Bianco2014}; (24) \citet{Stritzinger2012}; (25) \citet{Taddia2013}; (26) \citet{Modjaz2006, Pian2006, Sonbas2008}; (27) \citet{Valenti2008, Chen2014}; (28) \citet{Inserra2011}; (29) \citet{Inserra2013}; (30) \citet{Hicken2017}; (31) \citet{Sahu2009}; (32) \citet{Stritzinger2009}; (33) \citet{Pastorello2008, Taubenberger2011}; (34) \citet{Mazzali2008, Modjaz2009}; (35) \citet{Faran2014b}; (36) \citet{Roy2011}; (37) \citet{Pignata2009, Pignata2011}; (38) \citet{Inserra2012}; (39) \citet{Takats2015}; (40) \citet{Mauerhan2013a, Fraser2013, Margutti2014, Graham2014}; (41) \citet{Smartt2015}; (42) \citet{Childress2016}; (43) \citet{Valenti2011}; (44) \citet{Elias-Rosa2010}; (45) \citet{Takats2014}; (46) \citet{Pastorello2015}; (47) \citet{Valenti2012}; (48) \citet{Arcavi2011, Tsvetkov2012, Sahu2013, Ergon2014, Ergon2015}; (49) \citet{Milisavljevic2011, Milisavljevic2013a}; (50) \citet{Morales-Garoffolo2015, Kumar2013}; (51) \citet{Bufano2014}; (52) \citet{Roming2012, Humphreys2012, Mauerhan2013b}; (53) \citet{Karamehmetoglu2023}; (54) \citet{Quimby2018}; (55) \citet{Pian2020}; (56) \citet{Tomasella2013}; (57) \citet{Milisavljevic2015}; (58) \citet{Takaki2013, Milisavljevic2013b}; (59) \citet{Bayless2013, Bose2013, Munari2013, Dall'Ora2014}; (60) \citet{Rubin2016}; (61) \citet{Maund2013, Barbarino2015}; (62) \citet{Valenti2014b}; (63) \citet{Bose2015a, Silverman2017}; (64) \citet{Davis2021}; (65) \citet{Zhang2014, Khazov2016}; (66) \citet{Tomasella2018}; (67) \citet{Prentice2019}; (68) \citet{Valenti2015, Black2017}; (69) \citet{Silverman2012, VanDyk2014, Morales-Garoffolo2014}; (70) \citet{Yaron2012}; (71) \citet{Valenti2014, Bose2015b, Dhungana2016, Yuan2016}; (72) \citet{Kangas2016}; (73) \citet{Valenti2016, Yaron2017, Bullivant2018}; (74) \citet{Drout2016}; (75) \citet{Bose2016, Terreran2016}; (76) \citet{Taddia2016, Taddia2019b}; (77) \citet{Pastorello2018}; (78) \citet{Nakaoka2018, Hosseinzadeh2018}; (79) \citet{Prentice2018a, Terreran2019}; (80) \citet{Tartaglia2017}; (81) \citet{Reynolds2020}; (82) \citet{Huang2018}; (83) \citet{Kuncarayakti2018}; (84) \citet{Teffs2021}; (85) \citet{Anderson2018}; (86) \citet{Prentice2018b}; (87) \citet{Kankare2021}; (88) \citet{McBrien2019}; (89) \citet{Irani2022}; (90) \citet{Prentice2020, Chen2020}; (91) \citet{Yao2020}; (92) \citet{Yang2021}; (93) \citet{Valerin2022}; (94) \citet{Irani2024}; (95) \citet{Tinyanont2022}; (96) \citet{Sollerman2021}; (97) \citet{Terreran2022}; (98) \citet{Reguitti2022}; (99) \citet{Hosseinzadeh2022}

\end{minipage}}
\end{tabular}
\end{table*}


\bsp	
\label{lastpage}
\end{document}